\documentclass[12pt]{article}

\usepackage[margin=1in]{geometry} 
\usepackage{setspace}			

\usepackage{makeidx}
\usepackage{graphicx}
\graphicspath{{figures/}}

\usepackage{authblk}			
\usepackage{natbib}			
\usepackage{amsmath,amssymb}	
\usepackage{longtable}

\usepackage{booktabs}			
\usepackage{multirow}			
\usepackage{array}				

\usepackage{float}				
\usepackage[section]{placeins} 	
\usepackage{caption}			
\usepackage{subcaption}			

\usepackage{caption}
\usepackage{url}				

\usepackage{xcolor}				
\usepackage[colorlinks=true,linkcolor=blue,citecolor=blue,urlcolor=blue]{hyperref}
\usepackage[normalem]{ulem}
\usepackage{hyperref}
\usepackage[nameinlink,noabbrev]{cleveref}

\usepackage[T1]{fontenc}
\usepackage[utf8]{inputenc}
\usepackage{lmodern}

\usepackage{amsthm}

\begin{document}

\title{Portfolio Optimization for Commodity ETFs under Heavy-Tailed Returns}

\author[1]{Nicholas Appiah\thanks{Corresponding author,\texttt{niappiah@ttu.edu}}}
\author[2]{Ali Jaffri}
\author[1]{Dilmi C.W. Hettiachchi-Halpe-Kankanamalage}
\author[1]{Svetlozar T. Rachev}

\affil[1]{Department of Mathematics \& Statistics,Texas Tech University, Lubbock, TX, USA}
\affil[2]{College of Business, North Dakota State University, Fargo, ND 58102, USA}

\date{}
\maketitle

\begin{abstract}
This paper examines portfolio optimization for commodity exchange-traded funds (ETFs) under heavy-tailed return behavior. Using daily Bloomberg data for 30 U.S.-listed commodity ETFs from 12 December 2018 to 16 December 2024, we study funds spanning agriculture, energy, metals, and broad commodity index exposure. We compare a passive buy-and-hold portfolio with rolling-window optimized portfolios formed under mean--variance and conditional value-at-risk (CVaR) criteria, considering both long-only and restricted long--short strategies. The results showed substantial heterogeneity across commodity sectors, with energy and broad commodity index funds displaying pronounced volatility, skewness, and excess kurtosis. Historical optimization indicated that minimum-risk and CVaR-based portfolios provided more stable cumulative performance than tangent portfolios and generally improved Sharpe, Calmar, and STARR$_{0.95}$ ratios. Extreme-value diagnostics showed that optimized portfolios remained exposed to heavy downside tails, so improved risk-adjusted performance did not eliminate extreme-loss risk. A dynamic extension based on ARMA--GARCH marginal models, Student--$t$ copula dependence, and one-step-ahead predictive scenarios improved performance mainly when combined with minimum-risk or CVaR-based objectives. Dynamic mean--variance tangent portfolios performed less reliably, reflecting sensitivity to expected-return estimation error. Transaction-cost robustness checks further showed that the practical value of dynamic optimization depended on turnover control, with low-turnover dynamic CVaR tangent portfolios remaining more resilient to implementation costs. Overall, the analysis showed that commodity ETF allocation benefited most from conservative and downside-risk-aware optimization, while optimized portfolios continued to require explicit tail-risk and implementation diagnostics.
\end{abstract}


\section{Introduction}\label{sec:intro}

From an asset-allocation perspective, commodity exchange-traded funds (ETFs) provide a liquid and accessible way to obtain exposure to real assets. They can be traded in standard brokerage accounts and allow investors to participate in commodity markets without directly trading futures contracts or holding physical commodities. Commodity exposure is often viewed as useful because commodity prices are linked to inflation dynamics, supply shocks, geopolitical risk, weather conditions, and business-cycle fluctuations. These features suggest that commodity ETFs may improve diversification relative to portfolios invested only in equities and fixed-income securities.

Commodity ETFs, however, are not simply equity ETFs with different underlying assets. Many commodity ETFs obtain exposure through futures contracts. Their returns therefore reflect not only spot-price movements, but also futures term-structure effects, roll yield, collateral returns, liquidity conditions, and fund-specific implementation rules. These effects vary across agriculture, energy, metals, and broad commodity index products. As a result, commodity ETF returns can display strong volatility, skewness, excess kurtosis, drawdowns, and abrupt reversals. Portfolio construction in this setting therefore requires attention to both conventional risk and downside-tail risk.

The theoretical foundation for commodity futures returns is closely connected to storage, hedging pressure, and the futures basis. The theory of normal backwardation links futures risk premia to the compensation required by speculators for bearing hedging pressure from producers and consumers \citep{Keynes1930}. The theory of storage emphasizes the role of inventories, carrying costs, and convenience yields in the relationship between spot and futures prices \citep{Kaldor1939,Working1949,Brennan1958,Telser1958}. Empirical evidence further shows that the basis contains information about expected commodity futures returns and that convenience yields vary across commodities and through time \citep{FamaFrench1987}. These mechanisms matter for commodity ETFs because futures-based products inherit both commodity-market shocks and the return effects associated with rolling futures exposure.

A large empirical literature examines whether commodities improve portfolio allocation. Early evidence suggests that commodity futures can provide diversification benefits because they have historically displayed low correlation with stocks and bonds \citep{BodieRosansky1980}. Commodity index returns have also been interpreted as compensation for collateral returns, roll returns, and rebalancing across heterogeneous contracts \citep{Greer2000}. Using long historical samples, \citet{GortonRouwenhorst2006} show that diversified commodity futures portfolios can have attractive reward-to-risk properties and positive inflation sensitivity. At the same time, the benefits of commodity allocation are not uniform. Commodity returns depend on term-structure effects, hedging pressure, systematic risk, and estimation uncertainty \citep{Bessembinder1992,ErbHarvey2006,DaskalakiSkiadopoulos2011}. This mixed evidence motivates a portfolio-level evaluation of commodity ETFs rather than assuming that commodity exposure automatically improves portfolio performance.

The diversification role of commodities may also weaken when markets become more financially integrated. Index investment and the financialization of commodities have been associated with stronger cross-commodity comovement and tighter links between commodity futures and broader financial markets \citep{TangXiong2012}. Related work shows that financial participation can affect commodity prices, volatilities, and correlations through risk-sharing and information channels \citep{ChengXiong2014}. Commodity correlations also vary over time and may increase during periods of financial stress \citep{SilvennoinenThorp2013}. These findings are important for ETF investors because diversification benefits may be weakest precisely when downside protection is most needed.

These empirical features make commodity ETFs a useful setting for comparing variance-based and downside-risk-sensitive allocation rules. Mean--variance optimization remains the classical benchmark for portfolio construction because it provides a transparent framework for evaluating expected return relative to portfolio variance \citep{Markowitz1952}. However, variance is a symmetric measure of dispersion and does not distinguish upside variation from downside losses. This limitation is important in commodity markets, where returns often exhibit skewness, excess kurtosis, volatility clustering, and sharp reversals \citep{Cont2001}. Optimized portfolios are also sensitive to estimation error in means, variances, and covariances \citep{ChopraZiemba1993}. This sensitivity is especially relevant in rolling-window allocation, where weights are repeatedly estimated from finite samples. Constraints can reduce extreme weights and improve realized behavior, but they do not remove estimation risk \citep{JagannathanMa2003,DeMiguelGarlappiUppal2009,LedoitWolf2004}.

Conditional Value-at-Risk (CVaR) provides a natural complement to variance-based optimization. CVaR measures expected losses beyond a specified quantile threshold and therefore focuses directly on downside-tail exposure. It is also a coherent risk measure and can be implemented through tractable optimization formulations \citep{ArtznerEtAl1999,RockafellarUryasev2000}. For commodity ETF portfolios, CVaR is useful because it connects portfolio construction to severe-loss behavior rather than to average dispersion alone. Extreme value theory (EVT) provides an additional diagnostic framework for examining whether optimized portfolios remain exposed to heavy downside tails after diversification \citep{EmbrechtsKluppelbergMikosch1997}. In particular, Hill tail-index estimation allows the left tail of the optimized portfolio return distribution to be examined directly \citep{Hill1975,McNeilFrey2000}.

This paper contributes to this literature by studying commodity ETF allocation through an integrated portfolio-risk framework. Rather than evaluating commodity ETFs only as passive diversifiers or comparing optimized portfolios only through volatility-based performance measures, the paper combines historical rolling-window optimization, CVaR-based downside-risk control, dynamic ARMA--GARCH--Student--$t$ copula predictive scenarios, transaction-cost and turnover diagnostics, and EVT-based tail-risk analysis. This combination is important because commodity ETF portfolios may appear attractive under conventional reward-to-risk ratios while still retaining substantial exposure to extreme downside losses. The paper therefore asks not only whether optimization improves realized performance, but also whether the resulting portfolios remain vulnerable to heavy-tailed loss behavior.

This paper studies portfolio optimization for 30 U.S.-listed commodity ETFs from 12 December 2018 to 16 December 2024. The ETF universe spans agriculture, energy, metals, and broad commodity index funds. We compare a passive buy-and-hold portfolio with optimized portfolios formed under mean--variance and CVaR criteria. The optimization is implemented under both long-only and restricted long--short constraints. This design allows us to examine whether downside-risk-aware optimization improves realized performance and whether limited short exposure provides meaningful additional benefits.

The empirical analysis proceeds in two stages. The first stage uses historical rolling-window optimization. Portfolio weights are estimated from past returns and applied out of sample, producing realized portfolio return series for the buy-and-hold benchmark and the optimized strategies. The second stage introduces a dynamic extension. In that extension, ARMA--GARCH marginal models, Student--$t$ copula dependence, and one-step-ahead predictive scenarios are used to update portfolio weights over time. The purpose of the dynamic extension is not to compare all possible forecasting models, but to ask whether forward-looking distributional information improves portfolio construction when combined with mean--variance and CVaR objectives.

The paper makes four contributions. First, it provides a portfolio-level analysis of a representative universe of U.S.-listed commodity ETFs across agriculture, energy, metals, and broad commodity index funds, with attention to the heterogeneous volatility, skewness, excess kurtosis, and dependence patterns across sectors. Second, it compares mean--variance and CVaR allocation rules under both long-only and restricted long--short constraints, allowing the analysis to distinguish variance-based return seeking from downside-risk-aware portfolio construction. Third, it evaluates realized performance using cumulative value paths, Sharpe ratios, Calmar ratios, STARR ratios, turnover diagnostics, and transaction-cost robustness checks, thereby linking statistical performance to practical implementation. Fourth, it combines reward-to-risk measures with Hill tail-index diagnostics to assess whether optimized and dynamically rebalanced commodity ETF portfolios remain exposed to heavy downside tails. This design is important because stronger risk-adjusted performance does not necessarily imply that the left tail of the return distribution has become thin.

The results show that commodity ETF allocation benefits most from conservative and downside-risk-aware portfolio construction. Historical minimum-risk and CVaR-based portfolios generally improve risk-adjusted performance relative to the buy-and-hold benchmark, especially under Sharpe, Calmar, and STARR measures. The dynamic extension improves performance most clearly when paired with minimum-risk or CVaR-based objectives. In contrast, dynamic mean--variance tangent portfolios are less reliable because they are sensitive to expected-return estimation error. The EVT diagnostics show that optimization and dynamic rebalancing do not eliminate heavy downside-tail exposure. Overall, the evidence suggests that commodity ETF investors should combine portfolio optimization with explicit tail-risk diagnostics rather than relying on variance-based allocation alone.

The remainder of the paper is organized as follows. Section~\ref{sec:descriptive} describes the commodity ETF universe, data sources, market-capitalization profile, cumulative price behavior, summary statistics, and correlation structure. Section~\ref{sec:historical_results} presents the historical rolling-window portfolio optimization results. Section~\ref{sec:performance_measures} evaluates the optimized portfolios using reward-to-risk performance measures. Section~\ref{sec:evt_results} examines downside-tail behavior using extreme value diagnostics. Section~\ref{sec:dynamic_results} presents the dynamic portfolio optimization results based on predictive return scenarios. Section~\ref{sec:discussion} presents the discussion.

\section{Data and Descriptive Statistics}
\label{sec:descriptive}

This study uses daily data for 30 U.S.-listed commodity ETFs from 12 December 2018 to 16 December 2024. The ETF universe was selected to balance investability, sector coverage, and data consistency. Each fund had Bloomberg price, return, distribution, and market capitalization data available over the study period and contributed to one of the four commodity categories considered in the paper: agriculture, energy, metals, and broad commodity index funds. The resulting sample is therefore interpreted as a representative investable universe of commodity ETFs rather than the complete population of all commodity-linked exchange-traded products. This classification makes it possible to distinguish sector-specific exposures from diversified commodity-basket products.

Daily ETF prices, distributions, market capitalizations, and the U.S. three-month Treasury bill yield were obtained from Bloomberg Professional Services.\footnote{Accessed on 27 March 2026.} Adjusted daily prices were used to compute arithmetic returns. Arithmetic returns are used throughout the portfolio analysis because portfolio returns are formed as weighted sums of individual asset returns. The Treasury bill yield was converted to a daily rate and used as the short-term risk-free proxy. Table~\ref{tab:ETF} in Appendix~\ref{app:ETF} lists the ETFs by commodity category, ticker symbol, fund name, inception date, and market capitalization.

The sample displays substantial heterogeneity in both fund size and underlying commodity exposure. It includes three large-cap ETFs with market capitalization greater than or equal to \$10 billion, three mid-cap ETFs with market capitalization between \$2 billion and \$10 billion, ten small-cap ETFs with market capitalization between \$0.3 billion and \$2 billion, and fourteen ETFs with market capitalization below \$0.3 billion. This variation is important because fund size may affect liquidity, trading costs, investor adoption, and the practical interpretation of optimized portfolio weights.

\begin{figure}[h]
\includegraphics[width=\textwidth]{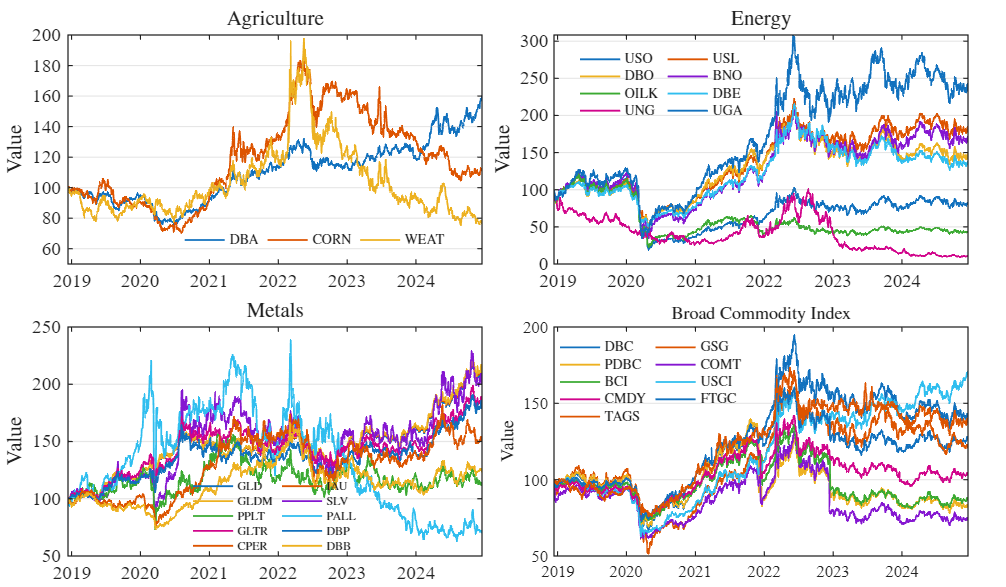}
\caption{Value of each commodity ETF under a buy-and-hold strategy. The values are normalized to 100 USD on 13 December 2018.}
\label{fig:indexed_prices}
\end{figure}

Figure~\ref{fig:indexed_prices} plots the value of a \$100 buy-and-hold investment in each commodity ETF over the sample period. Each series is normalized to 100 on 13 December 2018 so that funds with different price levels can be compared on a common scale. The figure reveals substantial cross-sector differences. Energy ETFs exhibit the largest price variation, reflecting the sensitivity of crude-oil, gasoline, and natural-gas funds to large commodity-market shocks. Metals ETFs display more heterogeneous behavior: gold-related funds have comparatively smoother paths, while silver, platinum, palladium, copper, and base-metal funds show stronger cyclical movements. Agricultural ETFs exhibit more moderate variation, although their paths still reflect commodity-specific shocks. Broad commodity index ETFs occupy an intermediate position because they combine exposures across several commodity groups.

Figure~\ref{fig:cum_returns} in Appendix~\ref{app:arithmetic} reports the corresponding cumulative arithmetic returns. These plots present the same buy-and-hold behavior in return form, with each ETF normalized to zero at the start of the sample. The cumulative-return evidence confirms the sectoral pattern shown in Figure~\ref{fig:indexed_prices}: energy funds display the largest swings, metals are mixed across precious and industrial metals, agriculture is more moderate but still episodic, and broad commodity index funds reflect the combined behavior of several commodity sectors.

Table~\ref{tab:summary_stats} in Appendix~\ref{app:stat} reports the sample mean, volatility, skewness, and excess kurtosis of daily arithmetic returns for each ETF. The summary statistics provide clear evidence of non-Gaussian return behavior. Energy ETFs have the highest dispersion, with natural-gas and crude-oil-related funds contributing some of the largest volatility values. Several energy funds also display large negative skewness and high excess kurtosis, indicating that energy ETF returns are strongly affected by abrupt downside shocks and volatility clustering. This pattern is consistent with the large cumulative swings observed in Figure~\ref{fig:indexed_prices}.

The broad commodity index funds also display important tail-risk features. Although these funds provide diversified exposure across commodity groups, several of them exhibit pronounced negative skewness and very large excess kurtosis. This result shows that diversification across commodity sectors does not eliminate heavy-tailed return behavior. Instead, broad commodity index products can still inherit extreme movements from energy, metals, and other commodity exposures, especially during periods of market stress or abrupt changes in futures-market conditions. Metals ETFs show mixed behavior, with lower volatility among gold-related funds and stronger tail behavior among silver, platinum, palladium, copper, and base-metal funds. Agricultural ETFs have more moderate volatility, but they still exhibit nonzero skewness and excess kurtosis. These distributional features motivate portfolio methods that account for downside risk and heavy-tailed returns rather than relying only on variance.

\begin{figure}[h!]
\centering
\includegraphics[width=0.9\linewidth]{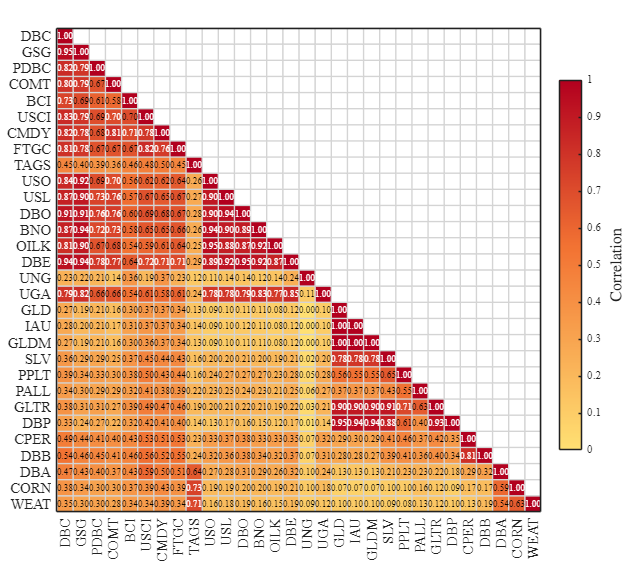}
\caption{Correlation heatmap of daily arithmetic returns for commodity ETFs, ordered by commodity category.}
\label{fig:corr_matrix}
\end{figure}

Figure~\ref{fig:corr_matrix} reports the correlation matrix of daily arithmetic returns. The ETFs are ordered by commodity category and, within each category, by increasing market capitalization. The heatmap shows clear sectoral dependence. Energy ETFs form the strongest within-sector block, reflecting common exposure to energy-market shocks. Metals ETFs form a separate dependence cluster, with stronger correlations among precious-metal funds and weaker links between metals and agriculture. Broad commodity index ETFs are connected to several sectors because their underlying exposures are diversified across commodity groups.

The descriptive evidence supports three conclusions. First, cumulative value paths differ substantially across commodity sectors, especially among energy, metals, agriculture, and broad commodity index funds. Second, daily return distributions exhibit skewness, excess kurtosis, and heavy-tail behavior, so variance alone provides an incomplete description of portfolio risk. This is especially important for energy and broad commodity index ETFs, where volatility and excess kurtosis are particularly pronounced. Third, the correlation matrix shows meaningful within-sector dependence, implying that diversification benefits depend on how the portfolio combines sector-specific and broad commodity exposures. These findings motivate the historical optimization, performance-ratio analysis, extreme value diagnostics, and dynamic allocation results that follow.

\section{Historical Portfolio Optimization}
\label{sec:historical_results}

Portfolio weights were first estimated using a historical rolling-window design. At each rebalancing date, the optimizer used the preceding 504 trading observations to estimate the relevant inputs, and the resulting weights were applied to the next trading day. This procedure generated out-of-sample optimized portfolio return series for the long-only and long--short strategies. The optimized portfolios were compared with a passive buy-and-hold portfolio (BHP), initialized with equal weights across the 30 commodity ETFs and held without rebalancing.

Let $r_t$ denote the vector of daily arithmetic ETF returns at time $t$, and let $w_t$ denote the portfolio weight vector selected before the return realization. Portfolio returns are computed as
\[
R_{p,t}=w_t^\top r_t .
\]
The full-investment condition requires portfolio weights to sum to one at each rebalancing date.

We considered two investment strategies: long-only and long--short. Under the long-only strategy, portfolio weights satisfy
\[
\sum_{i=1}^{30} w_i(t_k)=1,
\qquad
0 \leq w_i(t_k) \leq 1,
\]
where $w_i(t_k)$ denotes the weight assigned to ETF $i$ over the holding period $[t_k,t_{k+1})$. Under the long--short strategy, the bounds on the weights were relaxed to
\[
-w_s \leq w_i(t_k) \leq 1+w_s,
\qquad
w_s=\frac{1}{30}.
\]
The constant $w_s=1/30$ limits the short position in any single ETF to approximately the initial equal-weight allocation of the 30-ETF universe. This specification allowed limited short exposure while preventing any individual short position from becoming large relative to the portfolio.

Under both strategies, portfolio rebalancing was governed by the turnover constraint
\[
\frac{1}{2}\sum_{i=1}^{30}\left|w_i(t_k)-w_i(t_{k-1})\right| < C_{\mathrm{TO}}.
\]
The value $C_{\mathrm{TO}}=1/(30\times252)$ was imposed to keep rebalancing tightly controlled and to prevent the optimized portfolios from relying on excessive trading. This restriction is important because commodity ETFs can differ substantially in liquidity, bid--ask spreads, and implementation costs. Transaction costs are especially relevant for dynamically rebalanced portfolios because proportional trading costs penalize frequent changes in portfolio weights \citep{MagillConstantinides1976,Constantinides1986}. To evaluate practical implementation, we also report turnover diagnostics and proportional transaction-cost robustness checks in Appendix~\ref{app:turnover_allocation}. Let
\[
\mathrm{TO}_{p,t}
=
\frac{1}{2}\sum_{i=1}^{30}\left|w_i(t)-w_i(t-1)\right|
\]
denote the realized one-period portfolio turnover. For a proportional transaction-cost rate $c$, the transaction-cost-adjusted portfolio return is computed as
\[
R^{\mathrm{net}}_{p,t}
=
R_{p,t}
-
c\,\mathrm{TO}_{p,t}.
\]
Thus, strategies with higher realized turnover are penalized more heavily. The historical transaction-cost results are reported in Table~\ref{tab:tc_hist}, while the corresponding dynamic results are reported later in Table~\ref{tab:tc_dyn}.

We considered two optimization criteria. The first was the classical mean--variance framework of \citet{Markowitz1952}. The mean--variance minimum-risk portfolio is denoted MVP, while the corresponding tangent portfolio is denoted TVP. The second criterion was conditional value-at-risk (CVaR), following \citet{RockafellarUryasev2000}. CVaR optimization was implemented at the 95\% and 99\% confidence levels of the loss distribution, where losses are defined as negative portfolio returns. The minimum CVaR portfolios are denoted C95 and C99, and the corresponding CVaR tangent portfolios are denoted TC95 and TC99. The prefix LO denotes long-only portfolios, while LS denotes long--short portfolios.

For each strategy and optimization criterion, we considered two portfolio solutions on the frontier: the minimum-risk portfolio and the tangent portfolio. To distinguish among investment strategy, optimization method, and frontier solution, we use the following notation:
\begin{flushleft}
\begin{tabular}{ll}
\textbf{LO MVP} & long-only mean--variance minimum-variance portfolio; \\
\textbf{LO C95} & long-only CVaR(95\%) minimum-risk portfolio; \\
\textbf{LO C99} & long-only CVaR(99\%) minimum-risk portfolio; \\
\textbf{LO TVP} & long-only mean--variance tangent portfolio; \\
\textbf{LO TC95} & long-only CVaR(95\%) tangent portfolio; \\
\textbf{LO TC99} & long-only CVaR(99\%) tangent portfolio.
\end{tabular}
\end{flushleft}
The corresponding long--short portfolios are denoted analogously by replacing LO with LS. As an endogenous benchmark, we constructed a passive buy-and-hold portfolio (BHP), initialized with equal weights across the 30 commodity ETFs and held without rebalancing. The BHP weights were therefore allowed to evolve passively over time according to relative ETF performance.

\begin{figure}[h!]
\centering
\includegraphics[width=0.49\textwidth]{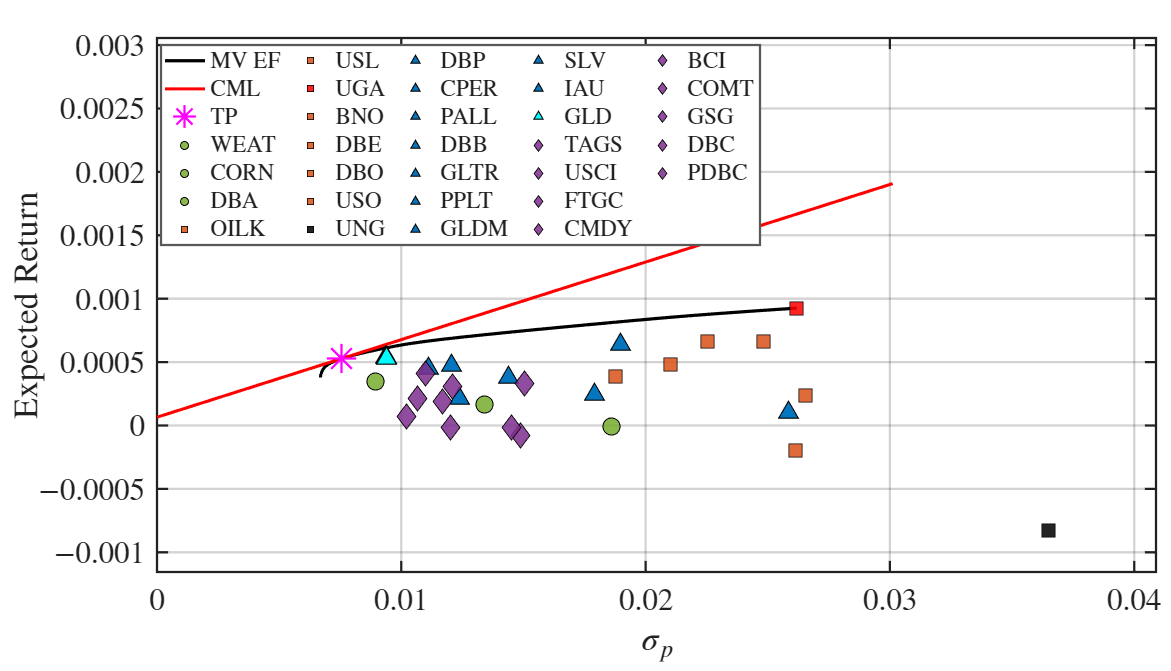}
\includegraphics[width=0.49\textwidth]{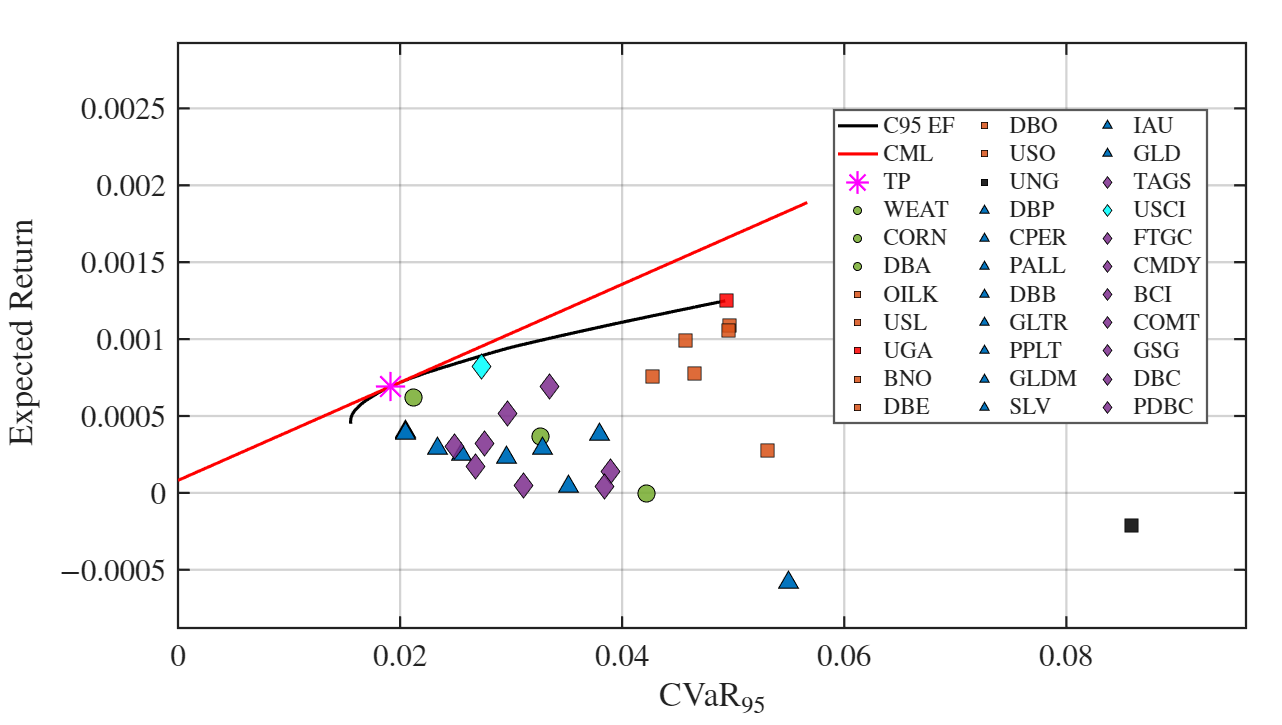}
\caption{Mean--variance efficient frontier (left) and CVaR95 efficient frontier (right) computed for 16 December 2024.}
\label{fig:EF}
\end{figure}

Figure~\ref{fig:EF} compares the historical mean--variance and CVaR efficient frontiers. For this frontier comparison, the inputs were estimated using the historical return window ending on 16 December 2024, the final date in the sample. The left panel reports the mean--variance frontier in expected-return--volatility space, while the right panel reports the CVaR$_{0.95}$ frontier in expected-return--tail-risk space. The capital market line is included in each panel, and the tangent portfolio is identified as the point on the corresponding frontier with the largest excess-return-to-risk ratio.

The two panels in Figure~\ref{fig:EF} provide complementary views of the commodity ETF opportunity set. The mean--variance frontier summarizes the trade-off between expected return and total return variability. The CVaR frontier instead summarizes the trade-off between expected return and estimated lower-tail loss. This distinction matters because the descriptive statistics in Section~\ref{sec:descriptive} show that several commodity ETFs, especially energy and broad commodity index funds, display skewness and excess kurtosis. A portfolio that appears attractive under volatility-based risk may not be equally attractive when risk is measured by expected downside loss.

\begin{figure}[h!]
\centering
\includegraphics[width=0.9\linewidth]{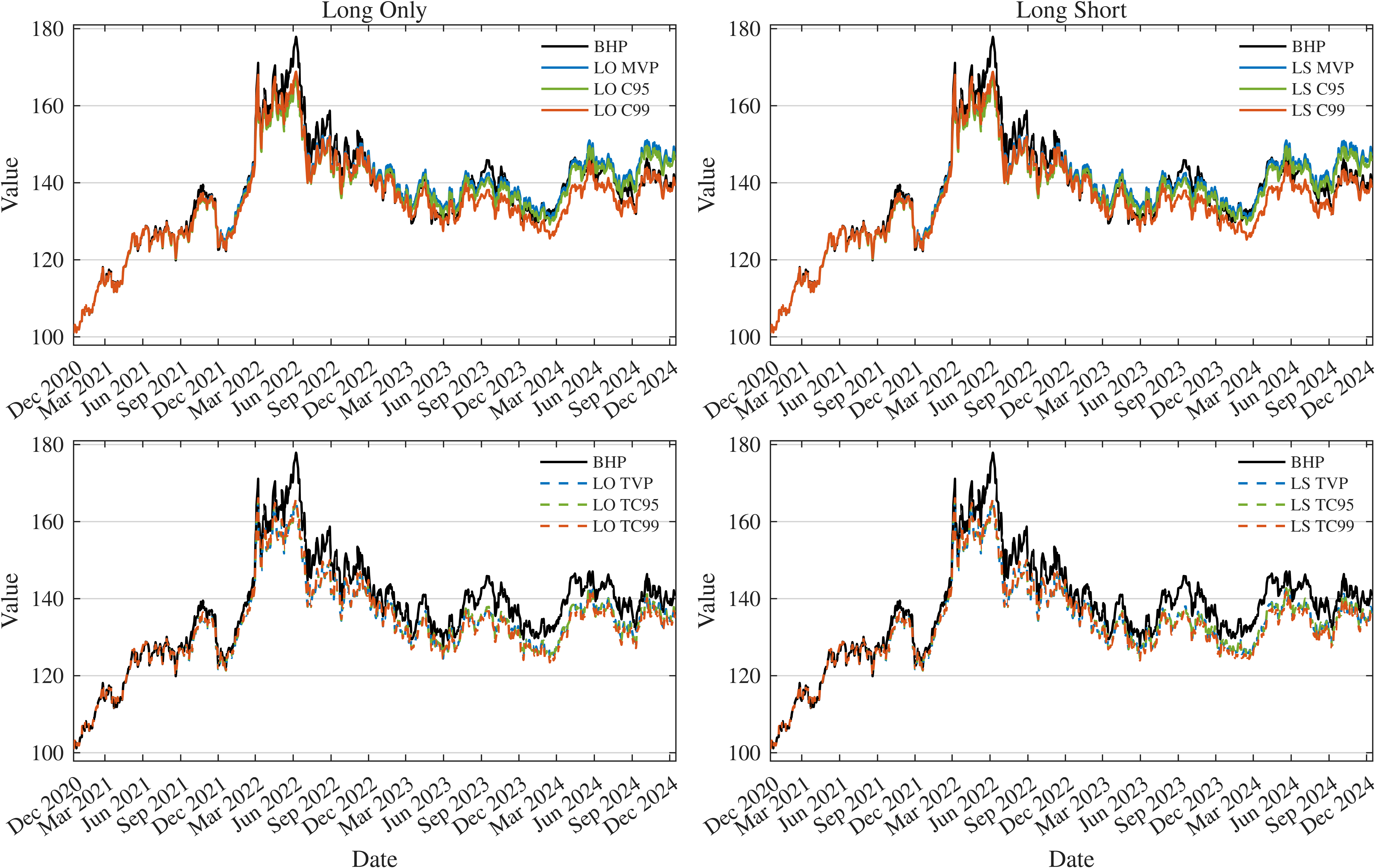}
\caption{Cumulative value of a \$100 investment under historical long-only and long--short optimized strategies.}
\label{fig:lo_ls}
\end{figure}

Figure~\ref{fig:lo_ls} reports the cumulative value of a \$100 investment under the historical rolling-window optimized strategies. The upper panels compare the minimum-risk portfolios, while the lower panels compare the tangent portfolios. The BHP is included in each panel as a passive benchmark.

The optimized portfolios and the BHP follow the same broad commodity-market cycle. Portfolio values rise during the 2021--2022 commodity rally, decline after the 2022 peak, and partially recover during 2024. The BHP reflects passive exposure because its weights drift with relative ETF performance. By contrast, the optimized portfolios are reweighted through the rolling-window procedure and therefore respond to changing return, volatility, and dependence conditions.

The minimum-risk portfolios in the upper panels display relatively stable cumulative value paths. The MVP, C95, and C99 strategies remain close to one another for much of the sample, with only modest separation around the 2022 peak and during the subsequent recovery. This indicates that variance-based and CVaR-based minimum-risk allocation rules identified broadly similar conservative regions of the commodity ETF opportunity set in the historical rolling-window setting.

The tangent portfolios in the lower panels behave differently. Although they follow the same broad commodity-market cycle, they generally produce weaker cumulative values than the minimum-risk portfolios after the 2022 peak. This suggests that return-seeking tangent allocation was more sensitive to estimation error and changing commodity-market conditions, consistent with the known sensitivity of optimized mean--variance portfolios to input-estimation error \citep{Michaud1989,ChopraZiemba1993,DeMiguelGarlappiUppal2009}. In this sample, the historical tangent portfolios did not deliver enough additional upside to compensate for their greater instability.

The long-only and long--short results are also visually close. Allowing short positions expanded the feasible set, but it did not automatically generate materially better cumulative performance. This is important because long--short flexibility can improve cross-sectional positioning, but it may also increase sensitivity to estimation error, leverage, transaction costs, borrowing costs, and short-sale implementation frictions. The turnover constraint, together with the transaction-cost robustness analysis in Appendix~\ref{app:turnover_allocation}, is therefore important for interpreting the practical relevance of the optimized strategies.

Overall, the historical optimization evidence shows that portfolio construction changed realized commodity ETF performance, but the benefits depended on the objective function. Minimum-risk and CVaR-based strategies provided more stable cumulative paths, while tangent portfolios were more fragile in the rolling-window setting. The results also suggest that implementation considerations should be evaluated jointly with portfolio performance because strategies with similar cumulative values may differ in turnover and transaction-cost sensitivity. The next section evaluates whether these cumulative-value patterns are confirmed by reward-to-risk performance measures.

\section{Performance Measures}
\label{sec:performance_measures}

To evaluate the historical optimized portfolios, we use three reward-to-risk performance measures: the Sharpe ratio, the Calmar ratio, and the stable tail-adjusted return ratio, denoted STARR. These measures complement the cumulative-value evidence in Section~\ref{sec:historical_results} by assessing whether higher portfolio values are achieved with better compensation for total volatility, maximum drawdown, and downside-tail losses.

For each portfolio $p$, let $R_{p,t}$ denote the daily portfolio return and let $r_{f,t}$ denote the daily risk-free rate. The daily excess return is
\[
Z_{p,t}=R_{p,t}-r_{f,t}.
\]
The Sharpe ratio is used as a standard excess-return-to-variability measure \citep{Sharpe1994}. It is computed as
\[
\mathrm{Sharpe}_p
=
\frac{\bar Z_p}{\sigma(Z_p)},
\]
where $\bar Z_p$ is the sample mean of daily excess returns and $\sigma(Z_p)$ is the sample standard deviation of daily excess returns.

The Calmar ratio is used to evaluate return performance relative to drawdown severity \citep{Young1991}. It is computed as
\[
\mathrm{Calmar}_p
=
\frac{\mathrm{CAGR}_p}{\mathrm{MDD}_p},
\]
where $\mathrm{CAGR}_p$ denotes the compound annual growth rate of portfolio $p$ and $\mathrm{MDD}_p$ denotes its maximum drawdown. The maximum drawdown is computed from the cumulative value path of the portfolio and measures the largest peak-to-trough decline over the evaluation period.

The STARR ratio is used to evaluate performance relative to downside-tail risk. For the daily excess return series $Z_{p,t}$, the STARR ratio at confidence level $\alpha$ is
\[
\mathrm{STARR}_{p,\alpha}
=
\frac{\bar Z_p}{\mathrm{CVaR}_{\alpha}(-Z_p)},
\]
where $\mathrm{CVaR}_{\alpha}(-Z_p)$ is the empirical conditional value-at-risk of excess losses \citep{ArtznerEtAl1999,RockafellarUryasev2000,RachevEtAl2008}. The reported STARR ratio is computed as $\mathrm{STARR}_{0.95}$, with CVaR evaluated at the 95\% confidence level for the daily excess-loss distribution.

\begin{figure}[h!]
\centering
\includegraphics[width=\linewidth]{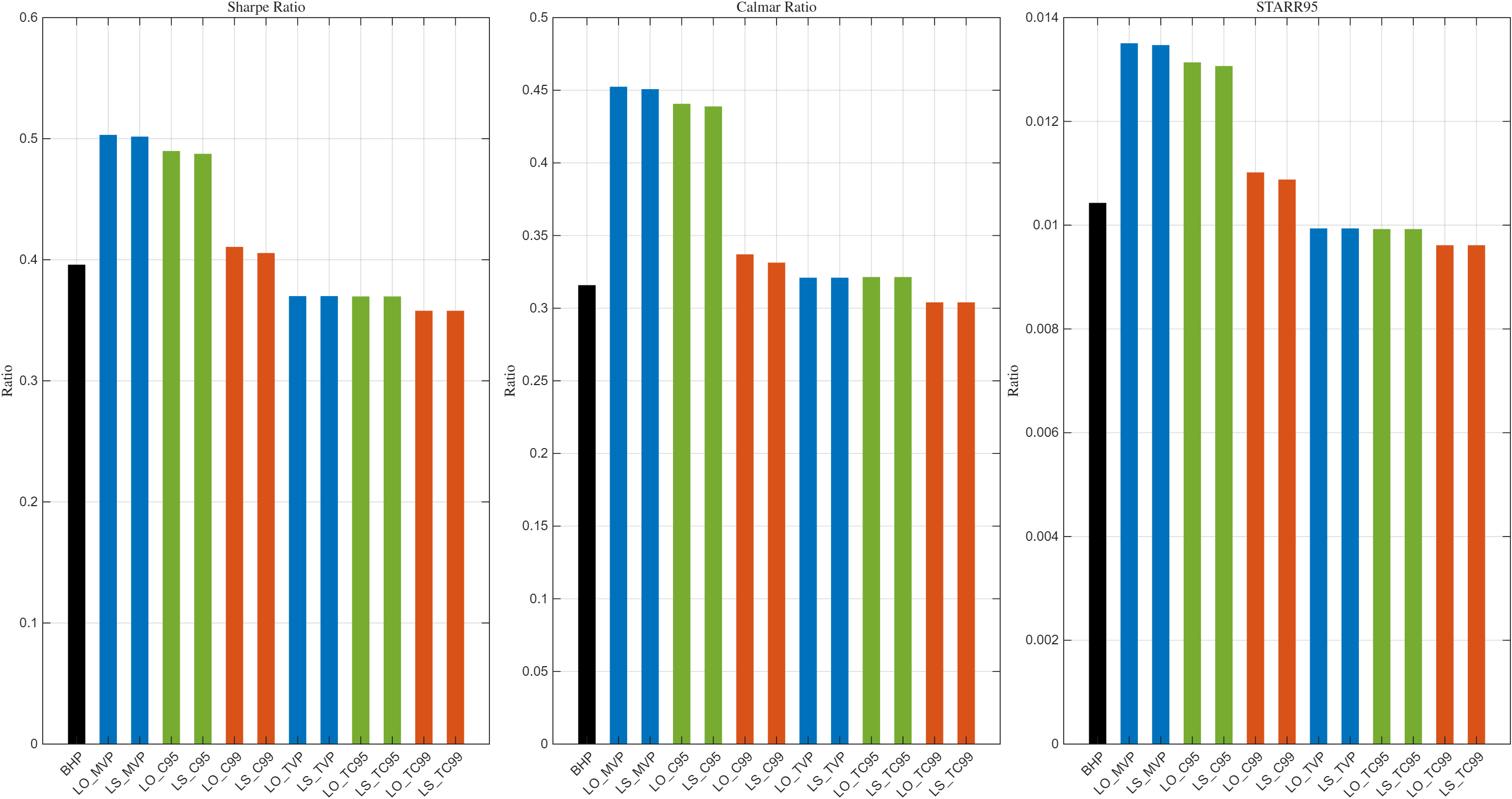}
\caption{Sharpe, Calmar, and STARR$_{0.95}$ ratios for the buy-and-hold portfolio and the optimized commodity ETF portfolios.}
\label{fig:performance_ratios}
\end{figure}

Figure~\ref{fig:performance_ratios} reports the Sharpe, Calmar, and STARR$_{0.95}$ ratios for the BHP and the historical optimized commodity ETF portfolios. The figure shows that the minimum-risk portfolios generally outperform the tangent portfolios across all three measures. The LO MVP and LS MVP portfolios produce the highest Sharpe ratios, followed closely by the LO C95 and LS C95 portfolios. The C99 portfolios produce weaker values than the MVP and C95 portfolios, but generally remain above the tangent portfolios.

The Calmar ratios provide a drawdown-based comparison. The MVP and C95 portfolios deliver stronger drawdown-adjusted performance than the BHP, the C99 portfolios, and the tangent portfolios. This indicates that the advantage of the minimum-risk portfolios is not driven only by lower return volatility, but also by better control of realized peak-to-trough losses. The BHP remains a useful passive benchmark, but it does not provide the same drawdown-adjusted performance as the leading minimum-risk strategies.

The STARR$_{0.95}$ results provide a tail-risk-adjusted comparison. The MVP and C95 portfolios generate the highest average excess return per unit of estimated CVaR$_{0.95}$ loss. This is important because commodity ETF returns exhibit skewness, excess kurtosis, and heavy-tail behavior, as shown in Section~\ref{sec:descriptive}. The STARR results therefore confirm that the leading minimum-risk portfolios perform well not only relative to total volatility and drawdown risk, but also relative to expected downside-tail losses.

Taken together, Figure~\ref{fig:performance_ratios} shows that the minimum-risk portfolios, especially MVP and C95, deliver the strongest historical reward-to-risk performance in this sample. Tangent portfolios perform less well across the three criteria, suggesting that return-seeking allocation is more fragile in the historical rolling-window setting. These results support the cumulative-value evidence in Section~\ref{sec:historical_results} and motivate the extreme value analysis in Section~\ref{sec:evt_results}, where we examine whether improved risk-adjusted performance also corresponds to thinner downside tails.

\section{Extreme Value Analysis}
\label{sec:evt_results}

The performance measures in Section~\ref{sec:performance_measures} show that the minimum-risk and CVaR-based portfolios provided stronger reward-to-risk performance than the buy-and-hold benchmark and the tangent portfolios. We now examine whether these gains are also reflected in the structure of extreme downside losses. This distinction is important because a portfolio may perform well according to Sharpe, Calmar, or STARR ratios while still remaining exposed to heavy left-tail risk.

Commodity ETF returns are especially suitable for extreme value analysis because large losses can arise from energy-market shocks, supply disruptions, geopolitical events, weather-related disturbances, and abrupt reversals in futures markets. Standard volatility-based risk measures may understate these risks when returns are skewed, leptokurtic, or clustered during turbulent periods. This concern is consistent with the descriptive evidence in Section~\ref{sec:descriptive}, where several energy and broad commodity index ETFs display pronounced skewness and excess kurtosis. Extreme value theory provides a direct way to evaluate the thickness of the tail of the loss distribution \citep{EmbrechtsKluppelbergMikosch1997,McNeilFrey2000}.

We focus on the left tail of the optimized portfolio return distribution. Let $R_{p,t}$ denote the return of portfolio $p$ at time $t$. The corresponding loss magnitude is defined as
\[
X_{p,t}
=
-R_{p,t}\mathbf{1}_{\{R_{p,t}<0\}}.
\]
Thus, $X_{p,t}$ is positive only when the portfolio return is negative and is equal to zero otherwise. Larger values of $X_{p,t}$ therefore correspond to larger downside losses. Let
\[
X_{(1)} \geq X_{(2)} \geq \cdots \geq X_{(m)}
\]
denote the ordered positive losses from largest to smallest. For a selected number $k$ of upper-order loss observations, the Hill estimator is
\[
\widehat{\alpha}(k)
=
\left[
\frac{1}{k}
\sum_{j=1}^{k}
\log\left(\frac{X_{(j)}}{X_{(k+1)}}\right)
\right]^{-1}.
\]
Lower values of $\widehat{\alpha}(k)$ indicate heavier downside tails, while higher values indicate thinner downside tails. Because the Hill estimator is sensitive to the threshold choice, we report Hill curves across a range of $k$ values rather than relying only on a single tail-index estimate.

\begin{figure}[h!]
\centering
\includegraphics[width=\linewidth]{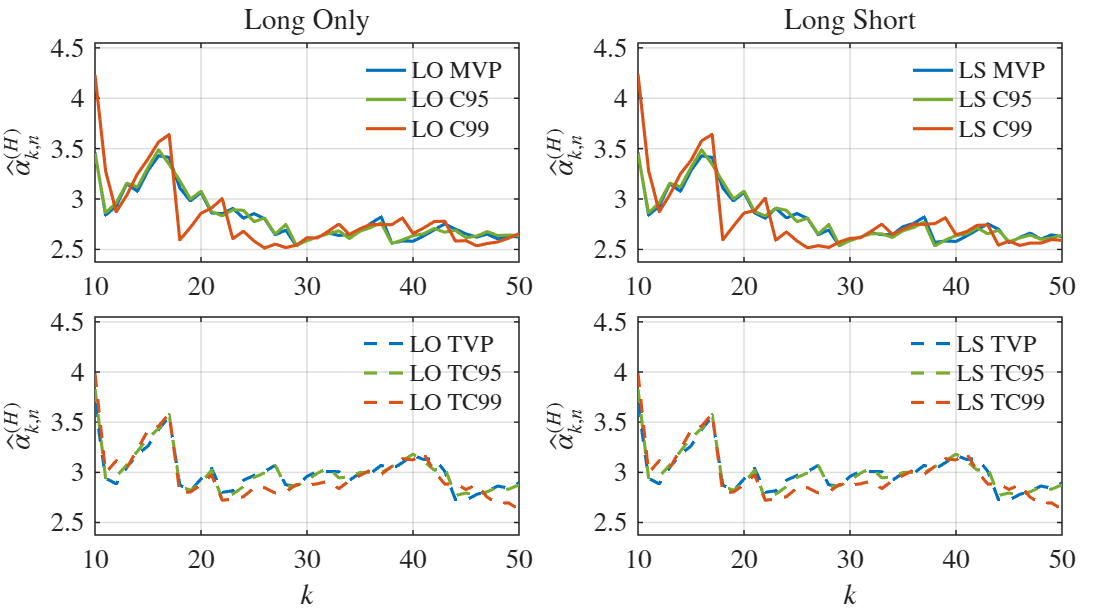}
\caption{Hill tail-index estimates for historical optimized commodity ETF portfolios using the largest 5\% of positive portfolio-loss observations. The upper panels report the minimum-risk portfolios, and the lower panels report the tangent portfolios. The left panels show long-only portfolios, and the right panels show long--short portfolios.}
\label{fig:hill_optimized}
\end{figure}

Figure~\ref{fig:hill_optimized} reports the Hill tail-index curves for the historical optimized commodity ETF portfolios. The upper panels compare the minimum-risk portfolios, while the lower panels compare the tangent portfolios. The left panels report long-only portfolios, and the right panels report long--short portfolios. The curves are computed from the largest 5\% of positive portfolio-loss observations, with very small threshold choices excluded to reduce instability.

The Hill estimates are more variable at smaller values of $k$, where the estimator depends on only a few extreme-loss observations. As $k$ increases, the curves become smoother and remain within a narrower range. This pattern is typical of tail-index diagnostics and reflects the standard bias--variance trade-off in extreme value estimation. Small values of $k$ focus on the most extreme losses but may be noisy, while larger values of $k$ provide more stable estimates but may include observations that are less clearly located in the extreme tail.

The minimum-risk portfolios in the upper panels exhibit relatively close Hill curves across both long-only and long--short implementations. This indicates that the MVP, C95, and C99 strategies had broadly similar estimated downside-tail behavior in the historical rolling-window setting. Although the CVaR-based portfolios directly target expected lower-tail losses, their Hill curves remain close to those of the mean--variance MVP portfolios. This suggests that CVaR optimization improved downside-risk-adjusted performance but did not eliminate heavy-tail exposure. In other words, CVaR control reduced the magnitude of realized tail losses, but it did not fundamentally remove the heavy-tailed structure of commodity ETF portfolio returns.

The tangent portfolios in the lower panels also display persistent downside-tail exposure. Their Hill curves are generally close to one another, but they show somewhat greater sensitivity across threshold choices. This is consistent with the performance evidence in Section~\ref{sec:performance_measures}, where tangent portfolios were weaker across volatility-, drawdown-, and tail-adjusted measures. The return-seeking nature of tangent allocation appears to make these portfolios more vulnerable to unstable tail behavior than the minimum-risk strategies.

The long-only and long--short panels are visually similar. This suggests that the restricted long--short specification did not materially reduce downside-tail exposure relative to the corresponding long-only strategies. Limited short exposure may expand the feasible allocation set, but it does not automatically thin the left tail of the optimized portfolio return distribution. This finding is important because long--short flexibility can improve allocation possibilities while still preserving exposure to extreme commodity-market losses.

To complement the Hill tail-index curves, Table~\ref{tab:tail_risk_hist} reports portfolio-level tail-risk measures for the historical optimized strategies. The table includes empirical VaR and CVaR at the 95\% and 99\% confidence levels, maximum drawdown, and the Hill tail-index estimate computed from the largest 5\% of positive loss observations.

\begin{table}[h!]
\centering
\caption{Historical tail-risk summary for optimized commodity ETF portfolios.}
\label{tab:tail_risk_hist}
\scriptsize
\begin{tabular}{lrrrrrr}
\toprule
Strategy & VaR$_{0.95}$ & CVaR$_{0.95}$ & VaR$_{0.99}$ & CVaR$_{0.99}$ & MDD & $\widehat{\alpha}$ \tabularnewline
\midrule
\multicolumn{7}{l}{\textbf{Minimum-risk portfolios}} \tabularnewline
\midrule
LO MVP  & 1.50 & 2.33 & 2.83 & 3.79 & 22.49 & 2.91 \tabularnewline
LS MVP  & 1.51 & 2.32 & 2.83 & 3.79 & 22.52 & 2.91 \tabularnewline
LO C95  & 1.50 & 2.32 & 2.82 & 3.79 & 22.53 & 2.89 \tabularnewline
LS C95  & 1.50 & 2.32 & 2.82 & 3.79 & 22.52 & 2.91 \tabularnewline
LO C99  & 1.53 & 2.37 & 3.04 & 3.88 & 25.66 & 2.61 \tabularnewline
LS C99  & 1.52 & 2.37 & 3.04 & 3.88 & 25.82 & 2.59 \tabularnewline
\midrule
\multicolumn{7}{l}{\textbf{Passive benchmark}} \tabularnewline
\midrule
BHP     & 1.78 & 2.76 & 3.22 & 4.43 & 28.34 & 2.78 \tabularnewline
\midrule
\multicolumn{7}{l}{\textbf{Tangent portfolios}} \tabularnewline
\midrule
LO TVP  & 1.63 & 2.44 & 2.96 & 3.90 & 25.03 & 2.82 \tabularnewline
LS TVP  & 1.63 & 2.44 & 2.96 & 3.90 & 25.03 & 2.82 \tabularnewline
LO TC95 & 1.62 & 2.43 & 2.98 & 3.89 & 24.96 & 2.78 \tabularnewline
LS TC95 & 1.62 & 2.43 & 2.98 & 3.89 & 24.96 & 2.78 \tabularnewline
LO TC99 & 1.58 & 2.42 & 3.00 & 3.89 & 25.66 & 2.73 \tabularnewline
LS TC99 & 1.58 & 2.42 & 3.00 & 3.89 & 25.66 & 2.73 \tabularnewline
\bottomrule
\end{tabular}

\vspace{2pt}
\noindent{\footnotesize Notes: VaR, CVaR, and MDD are reported in percent. VaR and CVaR are computed from daily portfolio losses. The Hill estimator is computed from the largest 5\% of positive loss observations. Lower Hill estimates indicate heavier downside tails.}
\end{table}

Table~\ref{tab:tail_risk_hist} confirms that the historical minimum-risk portfolios reduce downside losses relative to the buy-and-hold portfolio. The MVP and C95 portfolios have lower VaR$_{0.95}$, CVaR$_{0.95}$, VaR$_{0.99}$, CVaR$_{0.99}$, and maximum drawdown values than the passive benchmark. This supports the performance-ratio evidence that conservative optimization improved downside-loss control. However, the Hill estimates remain finite and relatively close across strategies, indicating that historical optimization reduced the magnitude of realized losses but did not eliminate heavy-tail exposure. The C99 portfolios also illustrate that more conservative tail-focused optimization does not necessarily imply a uniformly thinner estimated tail, since the Hill estimator captures tail thickness rather than only the level of realized losses.

Overall, the extreme value evidence reinforces the main message from the historical performance analysis. Minimum-risk and CVaR-based portfolios improved risk-adjusted performance, but optimized commodity ETF portfolios remained exposed to heavy downside tails. Portfolio optimization changed the distribution of losses, but it did not make the return distribution Gaussian or eliminate extreme-loss risk. For this reason, commodity ETF portfolios should be evaluated using both reward-to-risk performance measures and explicit tail-risk diagnostics. The next section examines whether a dynamic allocation framework based on predictive return scenarios improves portfolio performance while preserving this tail-risk perspective.

\section{Dynamic Portfolio Optimization}
\label{sec:dynamic_results}

The historical results show that minimum-risk and CVaR-based portfolios provided more stable performance than tangent portfolios in the rolling-window setting. We now examine whether a forward-looking dynamic allocation framework improves portfolio construction. The purpose of this extension is to determine whether predictive information about conditional volatility and cross-sectional dependence improves performance relative to the purely historical rolling-window approach.

The dynamic procedure replaces the historical return window used in the optimization with one-step-ahead predictive return scenarios. For each ETF, an ARMA--GARCH marginal model is estimated within the rolling window to capture conditional mean and conditional volatility dynamics. The standardized innovations are then used to estimate a Student--$t$ copula, which captures cross-sectional dependence and tail dependence across commodity ETFs. Predictive return scenarios are generated from the fitted marginal and copula models, and the same mean--variance and CVaR optimization rules are applied to those simulated scenarios. Thus, the dynamic framework retains the same long-only and long--short strategy definitions as the historical analysis, but updates portfolio weights using a forward-looking predictive distribution.

For ETF $i$, the marginal return model is written as
\[
r_{i,t}
=
\mu_{i,t}
+
\varepsilon_{i,t},
\qquad
\varepsilon_{i,t}
=
\sigma_{i,t}z_{i,t},
\]
where $\mu_{i,t}$ is the conditional mean, $\sigma_{i,t}$ is the conditional volatility, and $z_{i,t}$ is the standardized innovation. The fitted marginal distributions are combined through the Student--$t$ copula to generate joint one-step-ahead return scenarios. These scenarios preserve time-varying marginal volatility while allowing simulated ETF returns to reflect cross-sectional dependence and tail dependence.

\begin{figure}[h!]
\centering
\includegraphics[width=\linewidth]{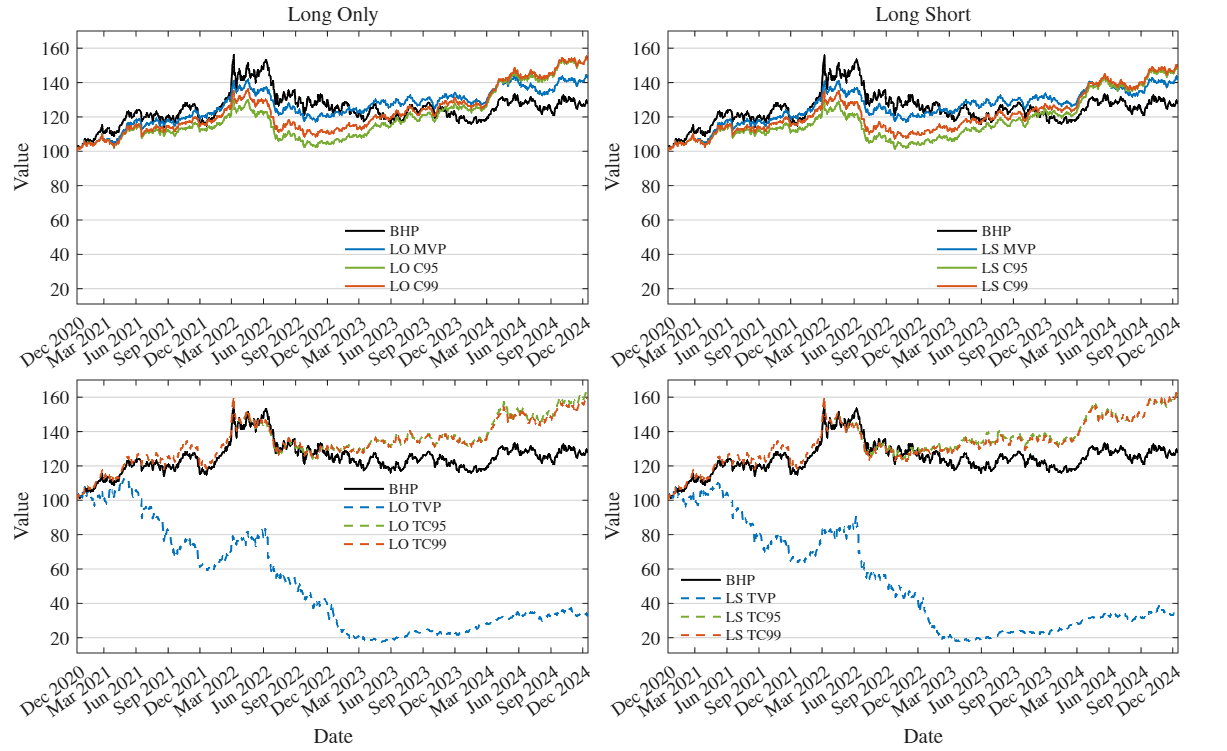}
\caption{Cumulative value of dynamically optimized commodity ETF portfolios. The upper panels report minimum-risk strategies, and the lower panels report tangent strategies. The left panels show long-only portfolios, and the right panels show long--short portfolios.}
\label{fig:dy_opt}
\end{figure}

Figure~\ref{fig:dy_opt} reports the cumulative value paths of the dynamically optimized portfolios. All strategies are initialized at 100, and the buy-and-hold portfolio is included as a passive benchmark. The upper panels compare the minimum-risk portfolios, while the lower panels compare the tangent portfolios. The left panels report long-only allocations, and the right panels report long--short allocations.

The dynamic minimum-risk portfolios display relatively stable cumulative value paths over the sample period. In both the long-only and long--short cases, the MVP, C95, and C99 strategies remain close to one another for much of the sample and generally stay above the BHP after the initial estimation period. This suggests that dynamic rebalancing based on a forward-looking return distribution improves portfolio stability relative to a passive allocation whose weights are allowed to drift over time. The CVaR-based minimum-risk portfolios are especially relevant because they target expected losses in the lower tail rather than total variability alone.

The tangent portfolios behave differently. In the lower panels of Figure~\ref{fig:dy_opt}, the dynamic TVP strategies experience substantially weaker cumulative performance than the CVaR-based tangent portfolios. This result is an important feature of the analysis rather than a minor empirical anomaly. Mean--variance tangent portfolios rely directly on conditional expected-return estimates, and these estimates are difficult to forecast accurately in heavy-tailed commodity ETF markets. Small errors in expected returns can translate into large changes in tangent-portfolio weights, especially when the optimization problem is solved repeatedly through time. The poor performance of the dynamic TVP portfolios therefore reflects the well-known sensitivity of return-seeking mean--variance allocation to expected-return estimation error.

By contrast, the dynamic TC95 and TC99 strategies remain much closer to the BHP and the minimum-risk portfolios. This shows that a downside-risk objective helps moderate the instability of tangent allocation. CVaR-based tangent portfolios still seek favorable reward-to-risk trade-offs, but they penalize expected losses beyond a tail threshold rather than relying only on volatility. In a commodity ETF universe characterized by skewness, excess kurtosis, and downside jumps, this tail-aware objective provides a more stable basis for dynamic allocation than variance-based tangency optimization alone.

The long--short results also show that allowing short positions does not automatically improve performance. Although short exposure expands the feasible allocation set, it may increase sensitivity to estimation error and tail dependence. The dynamic long--short TC95 and TC99 portfolios provide more stable performance than the long--short TVP, suggesting that downside-risk-sensitive optimization is especially important when the feasible set permits short positions. Overall, Figure~\ref{fig:dy_opt} shows that dynamic optimization is most effective when the predictive model is paired with minimum-risk or CVaR-based objectives rather than with variance-based tangency optimization alone.

\begin{figure}[h!]
\centering
\includegraphics[width=0.49\linewidth]{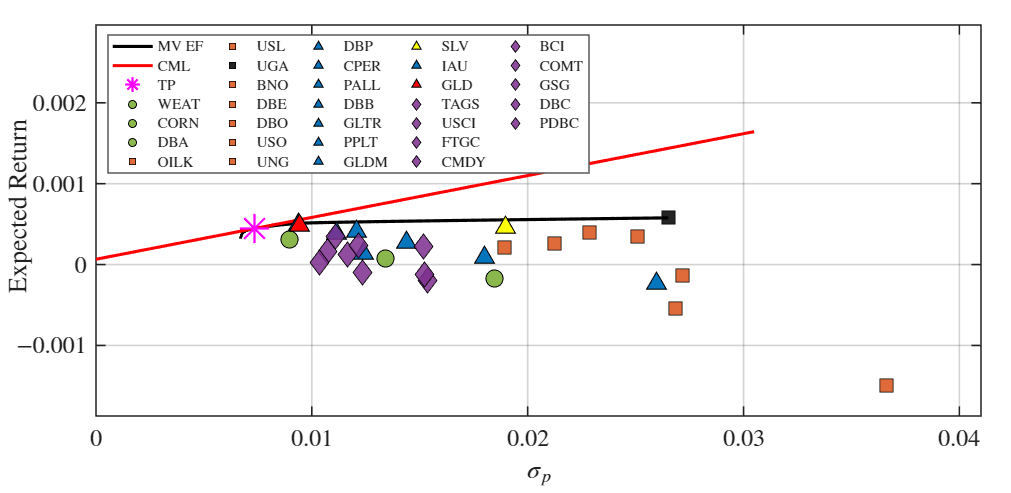}
\includegraphics[width=0.49\linewidth]{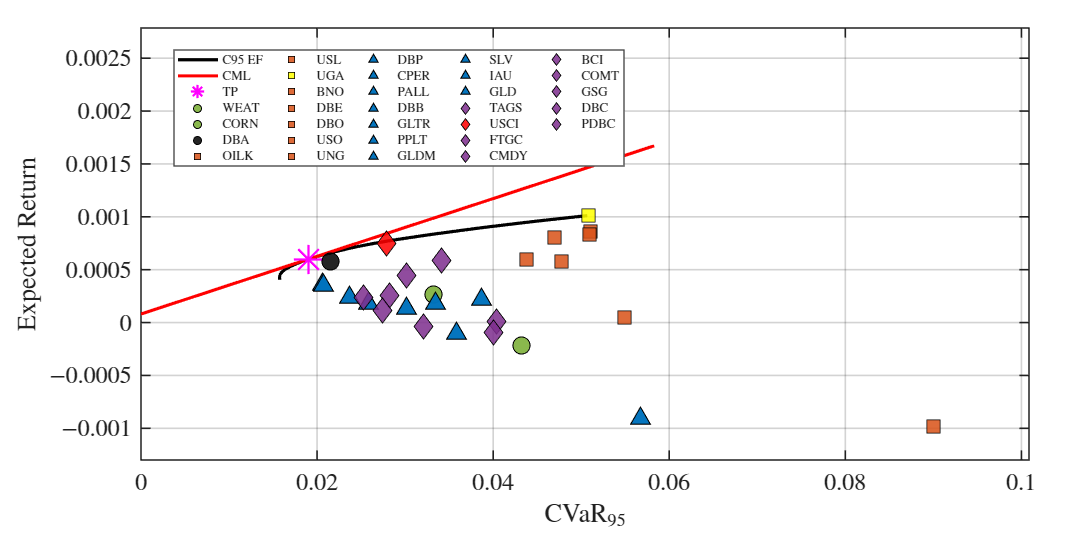}
\caption{Dynamic mean--variance and CVaR efficient frontiers constructed from ARMA--GARCH--copula predictive return scenarios.}
\label{fig:cvar_frontier}
\end{figure}

Figure~\ref{fig:cvar_frontier} reports the dynamic mean--variance and CVaR efficient frontiers constructed from ARMA--GARCH--copula predictive return scenarios. The left panel presents the mean--variance frontier in expected-return--volatility space, while the right panel presents the CVaR frontier in expected-return--tail-risk space.

The dynamic mean--variance frontier summarizes the conditional trade-off between expected return and portfolio standard deviation. Several individual ETFs lie below the frontier, indicating that diversified portfolios can improve the conditional risk-return trade-off relative to standalone commodity exposures. However, the cumulative paths in Figure~\ref{fig:dy_opt} show that the dynamic mean--variance tangent portfolio can still perform poorly out of sample when expected-return forecasts are unstable.

The CVaR frontier provides a complementary view by replacing volatility with coherent downside risk. Because commodity ETF returns exhibit skewness, excess kurtosis, and downside-tail behavior, the CVaR frontier is more directly aligned with the loss behavior documented in Section~\ref{sec:evt_results}. The CVaR-based efficient set identifies portfolios that reduce expected losses beyond the selected confidence threshold rather than merely reducing total return variability. This distinction is important in the dynamic setting because volatility and tail risk can evolve differently through time.

\begin{figure}[h!]
\centering
\includegraphics[width=\linewidth]{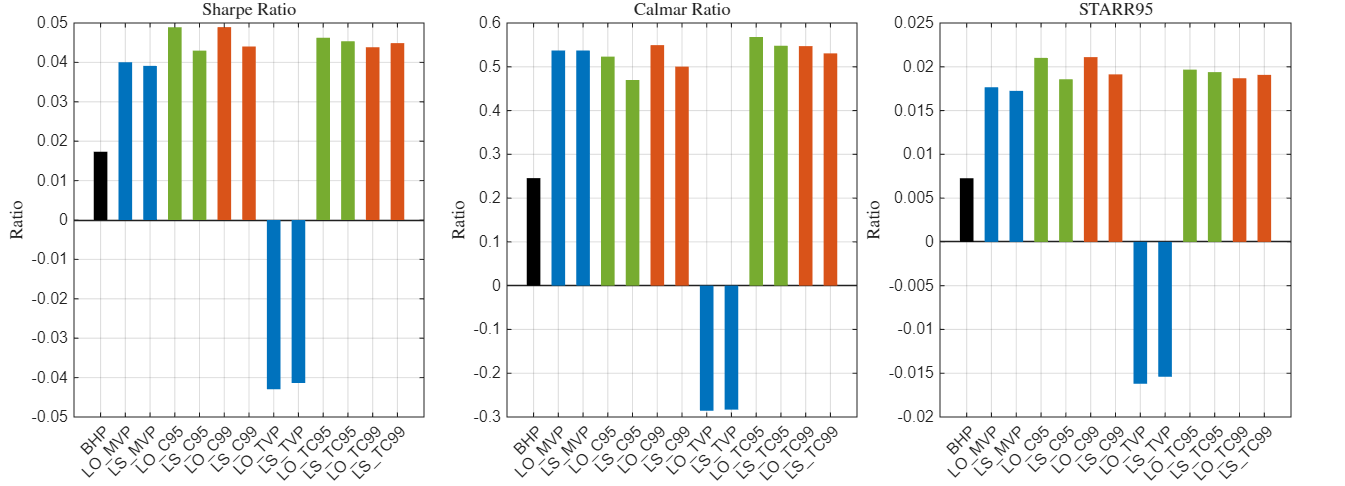}
\caption{Dynamic Sharpe, Calmar, and STARR$_{0.95}$ ratios for the buy-and-hold portfolio and dynamically optimized commodity ETF portfolios.}
\label{fig:dyn_ratio}
\end{figure}

Figure~\ref{fig:dyn_ratio} reports the Sharpe, Calmar, and STARR$_{0.95}$ ratios for the buy-and-hold portfolio and the dynamically optimized strategies. These metrics provide a direct comparison of dynamic portfolio performance across volatility-, drawdown-, and tail-risk-adjusted criteria.

The dynamic risk-ratio results show that performance improves most clearly for the minimum-risk and CVaR-based portfolios. The MVP, C95, and C99 strategies generally produce stronger Sharpe ratios than the BHP, indicating better excess return per unit of realized volatility. The Calmar ratios show that the same portfolios also provide stronger compensation for drawdown risk. This is important in commodity ETF markets, where volatility clustering and abrupt reversals can produce large interim losses. The STARR$_{0.95}$ results further show that CVaR-based dynamic strategies perform well when risk is measured by expected tail loss.

The dynamic TVP strategies behave very differently. Both long-only and long--short TVP allocations show weak performance across the three metrics. This pattern is consistent with the cumulative value paths in Figure~\ref{fig:dy_opt}, where the TVP portfolios deteriorate relative to the minimum-risk and CVaR-based strategies. The evidence suggests that conditional tangency optimization is highly sensitive to expected-return estimation error, especially in a heavy-tailed commodity ETF universe. This result is practically important because it shows that adding a sophisticated predictive model does not by itself solve the instability of return-seeking optimization. The objective function remains central: predictive scenarios are most useful when they are combined with conservative or downside-risk-aware allocation rules.

Table~\ref{tab:hist_dyn_compare} compares the historical and dynamic reward-to-risk performance of the optimized portfolios. The final three columns report dynamic-minus-historical differences. Statistical significance is assessed using moving-block bootstrap confidence intervals, which preserve short-run dependence in portfolio returns.

\begin{table}[htbp]
\centering
\caption{Historical and dynamic risk-adjusted performance comparison.}
\label{tab:hist_dyn_compare}
\scriptsize
\resizebox{\textwidth}{!}{%
\begin{tabular}{lrrrrrrrrr}
\toprule
& \multicolumn{3}{c}{Historical}
& \multicolumn{3}{c}{Dynamic}
& \multicolumn{3}{c}{Dynamic minus Historical} \tabularnewline
\cmidrule(lr){2-4}
\cmidrule(lr){5-7}
\cmidrule(lr){8-10}
Strategy
& Sharpe & Calmar & STARR$_{0.95}$
& Sharpe & Calmar & STARR$_{0.95}$
& $\Delta$ Sharpe & $\Delta$ Calmar & $\Delta$ STARR$_{0.95}$ \tabularnewline
\midrule
\multicolumn{10}{l}{\textbf{Minimum-risk portfolios}} \tabularnewline
\midrule
LO MVP  & 0.0353 & 0.4524 & 0.0151 & 0.0457 & 0.5373 & 0.0202 & 0.0104 & 0.0849 & 0.0051 \tabularnewline
LS MVP  & 0.0352 & 0.4508 & 0.0150 & 0.0447 & 0.5371 & 0.0198 & 0.0095 & 0.0863 & 0.0047 \tabularnewline
LO C95  & 0.0345 & 0.4407 & 0.0147 & 0.0543 & 0.5231 & 0.0234 & 0.0198 & 0.0824 & 0.0087 \tabularnewline
LS C95  & 0.0344 & 0.4389 & 0.0146 & 0.0484 & 0.4697 & 0.0209 & 0.0140 & 0.0309 & 0.0063 \tabularnewline
LO C99  & 0.0294 & 0.3370 & 0.0125 & 0.0543 & 0.5495 & 0.0235 & 0.0249 & 0.2124 & 0.0109 \tabularnewline
LS C99  & 0.0291 & 0.3313 & 0.0124 & 0.0494 & 0.5004 & 0.0215 & 0.0203 & 0.1691 & 0.0091 \tabularnewline
\midrule
\multicolumn{10}{l}{\textbf{Passive benchmark}} \tabularnewline
\midrule
BHP     & 0.0281 & 0.3159 & 0.0117 & 0.0207 & 0.2455 & 0.0087 & -0.0073 & -0.0703 & -0.0030 \tabularnewline
\midrule
\multicolumn{10}{l}{\textbf{Tangent portfolios}} \tabularnewline
\midrule
LO TVP  & 0.0268 & 0.3210 & 0.0114 & -0.0413 & -0.2857 & -0.0156 & -0.0681$^{*}$ & -0.6067$^{*}$ & -0.0270$^{*}$ \tabularnewline
LS TVP  & 0.0268 & 0.3210 & 0.0114 & -0.0398 & -0.2831 & -0.0148 & -0.0665$^{*}$ & -0.6041$^{*}$ & -0.0262$^{*}$ \tabularnewline
LO TC95 & 0.0268 & 0.3215 & 0.0114 & 0.0504 & 0.5680 & 0.0214 & 0.0236 & 0.2465 & 0.0100 \tabularnewline
LS TC95 & 0.0268 & 0.3215 & 0.0114 & 0.0495 & 0.5481 & 0.0212 & 0.0228 & 0.2266 & 0.0098 \tabularnewline
LO TC99 & 0.0260 & 0.3040 & 0.0111 & 0.0480 & 0.5472 & 0.0205 & 0.0220 & 0.2432 & 0.0094 \tabularnewline
LS TC99 & 0.0260 & 0.3040 & 0.0111 & 0.0490 & 0.5305 & 0.0209 & 0.0230 & 0.2265 & 0.0097 \tabularnewline
\bottomrule
\end{tabular}%
}
\vspace{2pt}

\noindent{\footnotesize Notes: The table reports Sharpe, Calmar, and STARR$_{0.95}$ ratios for historical and dynamic portfolio strategies. The final three columns report dynamic-minus-historical differences. $^{*}$ indicates that the 95\% moving-block bootstrap confidence interval for the corresponding difference excludes zero.}
\end{table}

Table~\ref{tab:hist_dyn_compare} shows that dynamic optimization improves the reward-to-risk measures for most minimum-risk and CVaR-based portfolios. The largest economic gains occur for the C99, TC95, and TC99 strategies, especially in the Calmar and STARR$_{0.95}$ ratios. However, the bootstrap evidence indicates that these positive improvements should be interpreted cautiously because their confidence intervals generally include zero. The statistically clearest result is the deterioration of the dynamic TVP portfolios. Both LO TVP and LS TVP have significantly negative dynamic-minus-historical differences across Sharpe, Calmar, and STARR$_{0.95}$. This confirms that dynamic mean--variance tangency optimization is fragile in the commodity ETF setting, while dynamic modeling is more useful when combined with conservative or downside-risk-aware objectives.

The comparison also shows that dynamic modeling does not automatically improve every strategy. The BHP has weaker performance over the dynamic comparison period, and the dynamic TVP portfolios perform substantially worse than their historical counterparts. By contrast, the dynamic TC95 and TC99 portfolios show strong economic improvements over their historical versions. This indicates that the dynamic framework is more reliable when the tangency objective is combined with CVaR-based downside-risk control.

The transaction-cost robustness results in Appendix~\ref{app:turnover_allocation} provide an additional implementation check. The historical optimized strategies are relatively insensitive to transaction costs because their realized turnover is tightly constrained. By contrast, the dynamic results in Table~\ref{tab:tc_dyn} show that high-turnover dynamic MVP, C95, C99, and TVP strategies can lose much of their gross performance advantage when trading costs are imposed. The dynamic TC95 and TC99 portfolios are more robust because their realized turnover is much lower. This distinction is important for practical implementation: dynamic allocation is most attractive when improved reward-to-risk performance is not driven by excessive rebalancing.

\begin{figure}[h!]
\centering
\includegraphics[width=\linewidth]{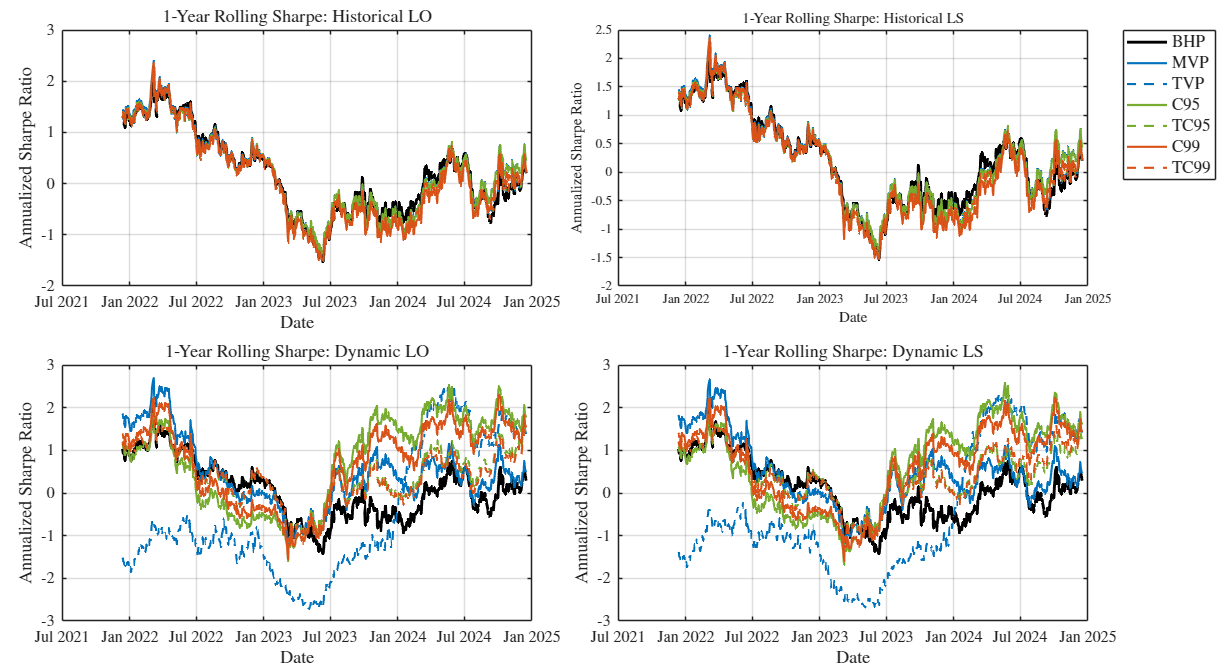}
\caption{One-year rolling Sharpe ratios for historical and dynamic commodity ETF portfolios. The upper panels report historical strategies, and the lower panels report dynamic strategies. The left panels show long-only portfolios, and the right panels show long--short portfolios.}
\label{fig:rolling_sharpe}
\end{figure}

To examine the time variation in reward-to-risk performance, Figure~\ref{fig:rolling_sharpe} reports one-year rolling Sharpe ratios for the historical and dynamic optimized portfolios. The upper panels show the historical strategies, while the lower panels show the dynamic strategies. The left panels report long-only portfolios, and the right panels report long--short portfolios.

Figure~\ref{fig:rolling_sharpe} shows that the historical long-only and long--short portfolios follow almost identical rolling Sharpe paths, indicating that the restricted long--short specification did not materially alter the historical reward-to-risk profile. The historical strategies rise during the 2021--2022 commodity rally, decline sharply through 2023, and partially recover in 2024. The dynamic panels show stronger separation across optimization objectives. In particular, the dynamic TVP portfolio performs poorly for much of the sample, while the CVaR-based dynamic strategies recover more strongly after 2023. This suggests that the dynamic framework is most effective when predictive modeling is combined with downside-risk-aware objectives rather than with mean--variance tangency optimization alone.

We also apply the Hill tail-index procedure to the dynamically optimized portfolio return series. This diagnostic evaluates whether dynamic modeling changes the estimated downside-tail behavior of the optimized commodity ETF portfolios.

\begin{figure}[h!]
\centering
\includegraphics[width=\linewidth]{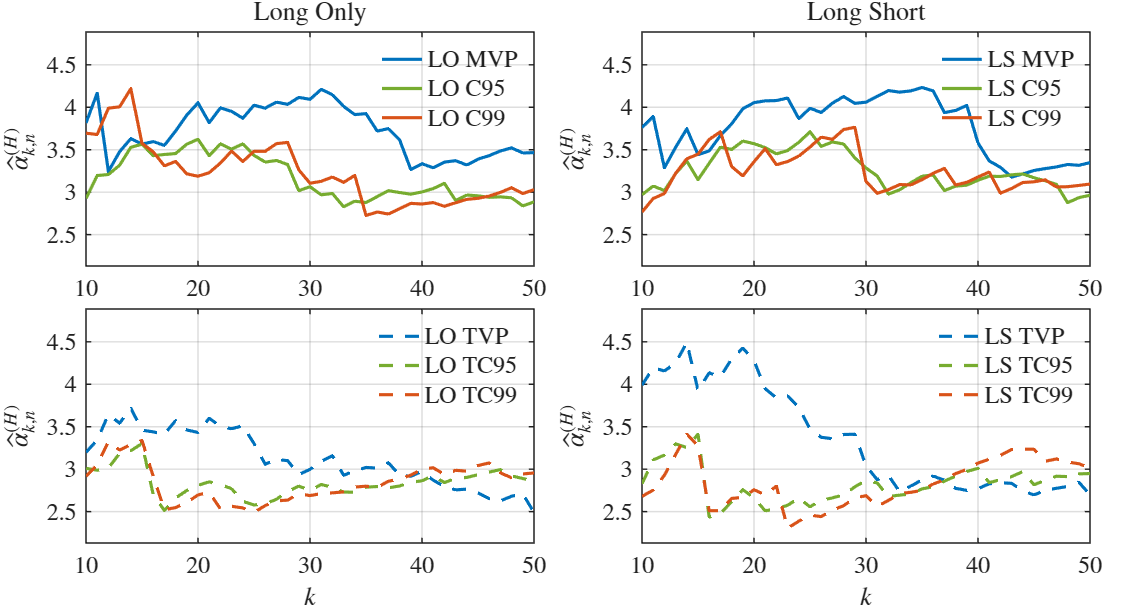}
\caption{Hill tail-index estimates for dynamically optimized commodity ETF portfolios. The upper panels report minimum-risk portfolios, and the lower panels report tangent portfolios. The left panels show long-only portfolios, and the right panels show long--short portfolios.}
\label{fig:dyn_hill}
\end{figure}

Figure~\ref{fig:dyn_hill} reports the Hill tail-index estimates for the dynamically optimized strategies. As in Section~\ref{sec:evt_results}, lower estimated tail exponents indicate heavier downside tails, while higher values indicate thinner downside tails. The estimates are more variable for smaller values of $k$, where the estimator uses only a few extreme-loss observations. For larger values of $k$, the curves become more stable and provide a clearer comparison across strategies.

The upper panels show the dynamic minimum-risk portfolios. The MVP portfolios generally produce tail-index estimates that are at least as high as those of the CVaR-based minimum-risk portfolios over much of the plotted range, indicating relatively thinner estimated downside tails. The C95 and C99 portfolios remain close to the MVP portfolios, suggesting that dynamic CVaR optimization improves downside-risk management without producing a sharply different tail-index profile. This distinction is important because CVaR controls expected losses beyond a quantile threshold, whereas the Hill estimator measures the relative thickness of the extreme-loss tail.

The lower panels report the dynamic tangent portfolios. The TVP, TC95, and TC99 strategies display more visible differences than the minimum-risk portfolios, especially for smaller values of $k$. These differences indicate that tangent strategies are more sensitive to the extreme observations used in the tail-index calculation. The CVaR-based tangent portfolios remain close to one another for larger values of $k$, suggesting that moving from the 95\% to the 99\% CVaR level changes the optimization problem but does not remove exposure to heavy downside tails.

Comparing the long-only and long--short panels shows that allowing short positions does not eliminate downside-tail exposure. Although long--short optimization expands the feasible set and can improve cross-sectional positioning, the resulting portfolios remain exposed to extreme commodity-market losses. This confirms that dynamic modeling improves adaptability but does not make the optimized return distribution Gaussian or free from tail risk.

Table~\ref{tab:tail_risk_dyn} reports the corresponding tail-risk measures for the dynamically optimized portfolios. This table complements the dynamic Hill curves by showing whether the dynamic strategies reduce realized loss magnitudes, drawdowns, and estimated tail thickness.

\begin{table}[htbp]
\centering
\caption{Dynamic tail-risk summary for optimized commodity ETF portfolios.}
\label{tab:tail_risk_dyn}
\scriptsize
\begin{tabular}{lrrrrrr}
\toprule
Strategy & VaR$_{0.95}$ & CVaR$_{0.95}$ & VaR$_{0.99}$ & CVaR$_{0.99}$ & MDD & Hill $\widehat{\alpha}$ \tabularnewline
\midrule
\multicolumn{7}{l}{\textbf{Minimum-risk portfolios}} \tabularnewline
\midrule
LO MVP  & 1.05 & 1.45 & 1.64 & 2.13 & 17.40 & 3.95 \tabularnewline
LS MVP  & 1.04 & 1.45 & 1.64 & 2.13 & 17.12 & 4.11 \tabularnewline
LO C95  & 1.04 & 1.54 & 1.70 & 2.42 & 21.49 & 3.51 \tabularnewline
LS C95  & 1.05 & 1.55 & 1.71 & 2.43 & 21.70 & 3.49 \tabularnewline
LO C99  & 1.06 & 1.55 & 1.81 & 2.38 & 20.55 & 3.34 \tabularnewline
LS C99  & 1.06 & 1.54 & 1.66 & 2.40 & 20.73 & 3.32 \tabularnewline
\midrule
\multicolumn{7}{l}{\textbf{Passive benchmark}} \tabularnewline
\midrule
BHP     & 1.69 & 2.52 & 2.82 & 4.14 & 25.93 & 2.91 \tabularnewline
\midrule
\multicolumn{7}{l}{\textbf{Tangent portfolios}} \tabularnewline
\midrule
LO TVP  & 3.78 & 5.89 & 6.72 & 9.15 & 84.91 & 3.51 \tabularnewline
LS TVP  & 3.99 & 6.04 & 7.04 & 9.15 & 84.38 & 3.73 \tabularnewline
LO TC95 & 1.36 & 2.05 & 2.39 & 3.43 & 22.94 & 2.82 \tabularnewline
LS TC95 & 1.35 & 2.02 & 2.32 & 3.39 & 23.19 & 2.53 \tabularnewline
LO TC99 & 1.36 & 2.04 & 2.36 & 3.41 & 22.63 & 2.53 \tabularnewline
LS TC99 & 1.38 & 2.05 & 2.32 & 3.44 & 23.84 & 2.80 \tabularnewline
\bottomrule
\end{tabular}

\vspace{2pt}
\noindent{\footnotesize Notes: VaR, CVaR, and MDD are reported in percent. VaR and CVaR are computed from daily portfolio losses. The Hill estimator is computed from the largest 5\% of positive loss observations. Lower Hill estimates indicate heavier downside tails.}
\end{table}

Table~\ref{tab:tail_risk_dyn} shows that the dynamic MVP, C95, and C99 portfolios reduce VaR and CVaR relative to the dynamic BHP. The dynamic MVP portfolios also have the lowest maximum drawdowns among the dynamic strategies and the highest Hill estimates, indicating thinner estimated downside tails. By contrast, the dynamic TVP portfolios have substantially larger VaR, CVaR, and maximum drawdown values, confirming the fragility of dynamic mean--variance tangency optimization. The dynamic TC95 and TC99 portfolios provide better downside-loss control than the dynamic TVP portfolios, but their Hill estimates show that they remain exposed to heavy-tail risk.

Overall, the dynamic results show that forward-looking ARMA--GARCH--Student--$t$ copula modeling improves commodity ETF portfolio construction mainly when it is paired with conservative or tail-aware allocation rules. Dynamic MVP, C95, C99, TC95, and TC99 portfolios generally show stronger economic performance than their historical counterparts across the main reward-to-risk measures. In contrast, dynamic mean--variance tangent portfolios are fragile because they depend heavily on noisy conditional expected-return estimates. The transaction-cost evidence further shows that dynamic performance must be interpreted jointly with turnover and implementation costs. The dynamic Hill diagnostics show that even successful dynamic strategies remain exposed to heavy downside tails. Dynamic allocation should therefore be interpreted as a tool for improving adaptability and downside-risk control, not as a way to eliminate extreme-loss risk.

\section{Discussion}
\label{sec:discussion}

This paper examined whether portfolio optimization improves commodity ETF allocation when returns are heavy-tailed, sector-dependent, and exposed to abrupt market reversals. The analysis combined descriptive return diagnostics, historical rolling-window optimization, reward-to-risk performance measures, extreme value analysis, transaction-cost robustness checks, and dynamic optimization based on predictive return scenarios. The central objective was to determine whether commodity ETF portfolios benefit more from variance-based return seeking or from conservative and downside-risk-aware allocation.

The descriptive evidence shows substantial heterogeneity across the commodity ETF universe. Energy ETFs display the largest cumulative swings and the highest volatility, reflecting their sensitivity to crude-oil, gasoline, and natural-gas market shocks. Metals ETFs show mixed behavior, with gold-related funds following comparatively smoother paths while silver, platinum, palladium, copper, and base-metal funds display stronger cyclical movements. Agricultural ETFs are less volatile than energy funds, but still exhibit episodic movements. Broad commodity index funds provide diversified exposure, but several of them still display pronounced negative skewness and large excess kurtosis. Thus, diversification across commodity sectors does not eliminate non-Gaussian return behavior or tail risk. This is an important empirical result because broad commodity ETF exposure may appear diversified in terms of holdings while still retaining substantial exposure to extreme downside movements.

The correlation evidence further shows that commodity ETFs do not behave as independent exposures. Energy funds form a strongly connected block, metals funds display their own dependence structure, and broad commodity index funds are linked to several sectors at once. This dependence structure matters for portfolio construction because diversification benefits depend not only on the number of funds held, but also on how sectoral exposures interact. A passive buy-and-hold allocation provides broad commodity exposure, but it does not actively respond to changes in volatility, dependence, or downside risk.

The historical optimization results show that portfolio construction materially affects realized performance. The minimum-risk portfolios provide more stable cumulative value paths than the tangent portfolios, while the buy-and-hold portfolio remains a useful passive benchmark. The MVP, C95, and C99 strategies remain close to one another in cumulative-value terms, suggesting that variance-based and CVaR-based conservative allocation rules identify similar stable regions of the commodity ETF opportunity set. By contrast, the tangent portfolios are more fragile. Their weaker cumulative performance after the 2022 commodity peak suggests that return-seeking optimization is sensitive to estimation error and changing commodity-market conditions.

The performance-ratio evidence strengthens this interpretation. Minimum-risk and CVaR-based portfolios generally outperform the buy-and-hold benchmark and the tangent portfolios across Sharpe, Calmar, and STARR$_{0.95}$ ratios. Their advantage is therefore not limited to one definition of risk. These strategies provide stronger compensation for total volatility, maximum drawdown, and expected downside-tail loss. The C95 portfolios are especially important because they combine downside-risk control with relatively strong realized performance. In this sample, moderate CVaR control appears more effective than aggressive return-seeking tangency optimization.

The extreme value analysis adds an important qualification. Although optimized portfolios improve risk-adjusted performance, the Hill tail-index diagnostics show that they remain exposed to heavy downside tails. This distinction is central to the paper. Better Sharpe, Calmar, or STARR ratios do not imply that extreme-loss risk has disappeared. Portfolio optimization changes the distribution of losses, but it does not make commodity ETF returns Gaussian or eliminate the possibility of large downside movements. Commodity ETF portfolios should therefore be evaluated using both reward-to-risk performance measures and explicit tail-risk diagnostics.

The dynamic results show that forward-looking modeling can improve portfolio construction when it is paired with robust allocation objectives. Dynamic minimum-risk and CVaR-based portfolios generally improve gross reward-to-risk performance relative to their historical counterparts. This suggests that updating portfolio weights using ARMA--GARCH--Student--$t$ copula predictive scenarios can be useful when the optimization objective is conservative or explicitly downside-risk-aware. The dynamic framework allows portfolio weights to respond to changes in conditional volatility and dependence, which are especially relevant in commodity markets.

However, dynamic modeling does not automatically improve performance. The dynamic mean--variance tangent portfolios perform poorly relative to the minimum-risk and CVaR-based strategies. This result highlights the danger of combining noisy conditional expected-return estimates with aggressive tangency optimization. In heavy-tailed commodity ETF markets, conditional mean forecasts can be unstable, and a strategy that maximizes expected excess return per unit of volatility may generate fragile allocations. The poor performance of the dynamic TVP portfolios is therefore an important finding: more sophisticated forecasting does not necessarily improve portfolio performance when the allocation rule is highly sensitive to expected-return estimation error.

The dynamic CVaR-based tangent portfolios provide a more constructive result. The TC95 and TC99 strategies perform substantially better than the dynamic TVP portfolios because the CVaR objective penalizes expected losses beyond a tail threshold. This makes the optimization less dependent on small differences in conditional expected returns and more directly aligned with the downside-loss behavior documented in the data. In heavy-tailed commodity ETF markets, this distinction is practically important. Dynamic allocation appears most useful when predictive scenarios are combined with objectives that control tail losses, rather than with mean--variance tangency optimization alone.

The transaction-cost robustness results further qualify the practical value of dynamic optimization. The historical optimized portfolios are relatively insensitive to transaction costs because their turnover is very low. In contrast, several dynamically optimized portfolios, especially the dynamic MVP, C95, C99, and TVP strategies, are more sensitive to proportional transaction costs because they require substantially more rebalancing. The dynamic TC95 and TC99 portfolios are more robust to trading frictions because their turnover is much lower. These results show that gross performance alone is not sufficient for evaluating commodity ETF strategies. Practical implementation requires joint consideration of performance, turnover, transaction costs, and allocation stability.

The long--short results also require careful interpretation. Allowing limited short positions expands the feasible set, but it does not uniformly improve performance or reduce tail risk. In several cases, the long-only and long--short portfolios behave similarly, reflecting the restrictive weight and turnover constraints. This is useful from an implementation perspective because it suggests that most of the gains in this sample come from the optimization objective rather than from shorting flexibility. At the same time, long--short strategies may still carry additional implementation concerns, including borrowing costs, short-sale constraints, liquidity frictions, and greater sensitivity to estimation error.

These findings have direct implications for institutional investors, commodity fund managers, pension funds, endowments, and family offices that use commodity ETFs for inflation hedging, diversification, real-asset exposure, or tactical allocation. The evidence suggests that commodity ETF portfolios should not be evaluated only by average return or volatility. Investors should also examine drawdown behavior, tail-loss exposure, turnover, transaction-cost sensitivity, and sector concentration. For institutions with long horizons and strict risk budgets, conservative and CVaR-based allocation rules may provide a more reliable framework than aggressive return-seeking tangency portfolios. For managers using dynamic allocation, the results suggest that forecast-based rebalancing should be paired with downside-risk controls and implementation constraints.

Overall, the evidence suggests that commodity ETF allocation should be built around downside-risk control rather than variance-based return chasing. Commodity ETFs provide liquid access to real-asset exposure, but they also carry sector-specific shocks, futures-market effects, volatility clustering, skewness, and heavy tails. A passive buy-and-hold portfolio offers broad exposure, but it does not actively manage these risks. Historical and dynamic optimization improve the allocation problem most clearly when they emphasize minimum-risk or CVaR-based objectives. The main implication is that commodity ETF portfolios should be evaluated jointly through cumulative performance, reward-to-risk ratios, turnover and transaction-cost diagnostics, allocation diagnostics, and explicit extreme value analysis.

\clearpage
\appendix
\counterwithin{figure}{section}
\counterwithin{table}{section}
\renewcommand{\thefigure}{\thesection\arabic{figure}}
\renewcommand{\thetable}{\thesection\arabic{table}}

\section{ETF Descriptions}
\label{app:ETF}

\begin{table}[h!]
\centering
\caption{Commodity ETF details by category, including inception dates and market capitalizations.}
\label{tab:ETF}
\begin{tabular}{llcc}
\toprule
\textbf{NYSE} & \textbf{\qquad Name} & \textbf{Inception} & \textbf{Market Cap} \\
\textbf{Ticker} & & \textbf{Date} & \textbf{(\$bn)} \\
\midrule

\multicolumn{4}{c}{\textbf{Agriculture}} \\
WEAT & Teucrium Wheat Fund & 09/19/2011 & 0.14 \\
CORN & Teucrium Corn Fund & 06/09/2010 & 0.17 \\
DBA  & Invesco DB Agriculture Fund & 01/05/2007 & 0.86 \\

\midrule
\multicolumn{4}{c}{\textbf{Energy}} \\
OILK & ProShares K-1 Free Crude Oil Strategy ETF & 11/24/2016 & 0.06 \\
USL  & United States 12 Month Oil Fund & 12/06/2007 & 0.09 \\
UGA  & United States Gasoline Fund & 02/26/2008 & 0.10 \\
BNO  & United States Brent Oil Fund & 06/02/2010 & 0.18 \\
DBE  & Invesco DB Energy Fund & 01/05/2007 & 0.20 \\
DBO  & Invesco DB Oil Fund & 01/05/2007 & 0.27 \\
USO  & United States Oil Fund & 04/10/2006 & 1.42 \\
UNG  & United States Natural Gas Fund & 04/18/2007 & 1.52 \\

\midrule
\multicolumn{4}{c}{\textbf{Metals}} \\
DBP  & Invesco DB Precious Metals Fund & 01/05/2007 & 0.14 \\
CPER & United States Copper Index Fund & 11/15/2011 & 0.29 \\
PALL & abrdn Physical Palladium Shares ETF & 01/08/2010 & 0.31 \\
DBB  & Invesco DB Base Metals Fund & 01/05/2007 & 0.36 \\
GLTR & abrdn Precious Metals Basket Shares ETF & 10/27/2010 & 0.37 \\
PPLT & abrdn Physical Platinum Shares ETF & 01/08/2010 & 0.97 \\
GLDM & SPDR Gold MiniShares Trust & 06/25/2018 & 7.84 \\
SLV  & iShares Silver Trust & 04/21/2006 & 11.47 \\
IAU  & iShares Gold Trust & 01/21/2005 & 31.12 \\
GLD  & SPDR Gold Shares & 11/18/2004 & 58.40 \\

\midrule
\multicolumn{4}{c}{\textbf{Broad Commodity Index}} \\
TAGS & Teucrium Agricultural Fund (Fund of Funds) & 03/28/2012 & 0.05 \\
USCI & United States Commodity Index Fund & 08/10/2010 & 0.12 \\
FTGC & First Trust Global Tactical Commodity Strategy Fund & 10/22/2013 & 0.21 \\
CMDY & iShares Bloomberg Roll Select Commodity Strategy ETF & 06/19/2018 & 0.24 \\
BCI  & abrdn Bloomberg All Commodity Strategy ETF & 03/30/2017 & 0.39 \\
COMT & iShares GSCI Commodity Dynamic Roll Strategy ETF & 10/15/2014 & 0.90 \\
GSG  & iShares S\&P GSCI Commodity Indexed Trust & 07/10/2006 & 1.23 \\
DBC  & Invesco DB Commodity Index Tracking Fund & 02/03/2006 & 2.33 \\
PDBC & Invesco Optimum Yield Diversified Commodity Strategy ETF & 11/07/2014 & 5.15 \\

\bottomrule
\end{tabular}
\end{table}

\section{Descriptive Statistics}
\label{app:stat}

\begin{table}[h!]
\centering
\caption{Summary statistics of daily arithmetic returns for commodity ETFs.}
\label{tab:summary_stats}
\begin{tabular}{lcccc}
\toprule
\textbf{Ticker} & $\boldsymbol{\mu}$ & $\boldsymbol{\sigma}$ & $\boldsymbol{\gamma}$ & $\boldsymbol{\kappa-3}$ \\
\midrule

\multicolumn{5}{c}{\textbf{Agriculture}} \\
WEAT & $-0.000005$ & 0.018590 & 0.792800 & 6.386100 \\
CORN & 0.000164 & 0.013415 & 0.127720 & 3.276900 \\
DBA  & 0.000349 & 0.008931 & $-0.396310$ & 2.199500 \\

\midrule
\multicolumn{5}{c}{\textbf{Energy}} \\
OILK & $-0.000195$ & 0.026142 & $-1.510200$ & 17.491000 \\
USL  & 0.000659 & 0.022551 & $-1.064200$ & 11.187000 \\
UGA  & 0.000926 & 0.026182 & $-0.658580$ & 12.472000 \\
BNO  & 0.000663 & 0.024845 & $-0.540570$ & 11.621000 \\
DBE  & 0.000384 & 0.018762 & $-0.770440$ & 5.388300 \\
DBO  & 0.000478 & 0.021022 & $-0.773930$ & 5.422600 \\
USO  & 0.000233 & 0.026538 & $-1.323700$ & 15.714000 \\
UNG  & $-0.000827$ & 0.036510 & 0.029156 & 1.173000 \\

\midrule
\multicolumn{5}{c}{\textbf{Metals}} \\
DBP  & 0.000450 & 0.011120 & $-0.121300$ & 4.203100 \\
CPER & 0.000378 & 0.014388 & $-0.068565$ & 1.723800 \\
PALL & 0.000100 & 0.025841 & $-0.098572$ & 6.805300 \\
DBB  & 0.000216 & 0.012382 & $-0.261510$ & 2.751100 \\
GLTR & 0.000476 & 0.012059 & $-0.351450$ & 4.230100 \\
PPLT & 0.000245 & 0.017921 & $-0.326940$ & 4.212600 \\
GLDM & 0.000540 & 0.009353 & $-0.288270$ & 2.608300 \\
SLV  & 0.000641 & 0.018954 & $-0.139680$ & 5.302300 \\
IAU  & 0.000535 & 0.009378 & $-0.297790$ & 2.438400 \\
GLD  & 0.000530 & 0.009406 & $-0.281440$ & 2.639900 \\

\midrule
\multicolumn{5}{c}{\textbf{Broad Commodity Index}} \\
TAGS & 0.000193 & 0.011680 & 0.199880 & 2.794600 \\
USCI & 0.000414 & 0.011000 & $-1.128700$ & 12.689000 \\
FTGC & 0.000214 & 0.010657 & $-1.054200$ & 16.503000 \\
CMDY & 0.000074 & 0.010197 & $-2.422100$ & 29.278000 \\
BCI  & $-0.000018$ & 0.012031 & $-3.700800$ & 46.986000 \\
COMT & $-0.000077$ & 0.014860 & $-3.761000$ & 44.913000 \\
GSG  & 0.000333 & 0.015061 & $-0.989980$ & 7.777800 \\
DBC  & 0.000304 & 0.012099 & $-0.795240$ & 4.414300 \\
PDBC & $-0.000014$ & 0.014517 & $-5.179500$ & 80.931000 \\

\bottomrule
\end{tabular}

\vspace{2pt}
\noindent{\footnotesize{
Here, $\mu$ denotes the sample mean, $\sigma$ denotes volatility measured by the sample standard deviation, 
$\gamma$ denotes skewness, and $\kappa-3$ denotes excess kurtosis.
}}
\end{table}

\section{ETF Cumulative Returns}
\label{app:arithmetic}

\begin{figure}[h!]
\centering
\includegraphics[width=\textwidth]{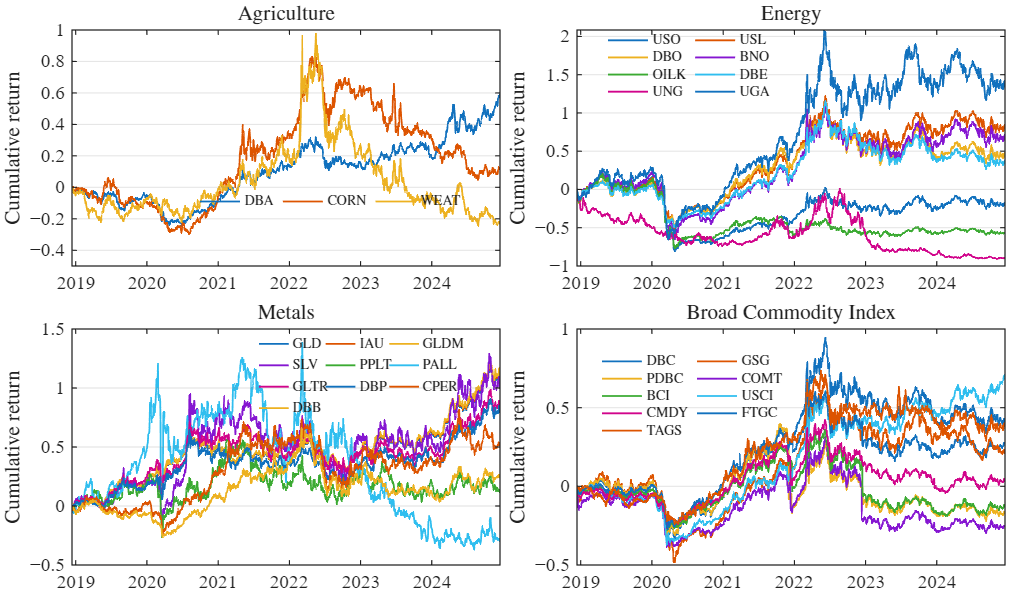}
\caption{Cumulative arithmetic returns of each commodity ETF under a buy-and-hold strategy. The returns were normalized to zero on 13 December 2018.}
\label{fig:cum_returns}
\end{figure}

Figure~\ref{fig:cum_returns} reports cumulative arithmetic returns for the individual commodity ETFs. The returns are normalized to zero at the beginning of the sample so that the cumulative return paths can be compared across funds with different price levels.

\section{Turnover, Transaction Costs, and Allocation Diagnostics}
\label{app:turnover_allocation}

This appendix reports implementation and allocation diagnostics for the historical and dynamic optimized portfolios. These diagnostics complement the main performance analysis by showing how frequently the portfolios are rebalanced, how sensitive the strategies are to proportional transaction costs, and how the optimized weights are distributed across commodity sectors and individual ETFs.

Table~\ref{tab:turnover_diagnostics} reports turnover and average exposure diagnostics. Turnover is computed as one-half of the sum of absolute changes in portfolio weights across consecutive rebalancing dates. The historical strategies exhibit very low turnover, while several dynamic strategies require substantially higher rebalancing. This indicates that the dynamic framework reacts more strongly to changes in the predictive return distribution. However, the dynamic TC95 and TC99 strategies show much lower turnover than the other dynamic strategies, suggesting that CVaR-based tangent allocation can be more stable when downside-risk control is imposed. The exposure diagnostics also show that the restricted long--short specification did not always generate meaningful short exposure, especially in the dynamic setting, where the optimized solutions were often effectively long-only.

\begin{table}[h!]
\centering
\caption{Historical and dynamic turnover and exposure diagnostics.}
\label{tab:turnover_diagnostics}
\resizebox{\textwidth}{!}{%
\begin{tabular}{lrrrrrrrrrr}
\toprule
& \multicolumn{5}{c}{Historical}
& \multicolumn{5}{c}{Dynamic} \tabularnewline
\cmidrule(lr){2-6}
\cmidrule(lr){7-11}
Strategy
& Daily Turn. & Ann. Turn. & Gross & Long & Short
& Daily Turn. & Ann. Turn. & Gross & Long & Short \tabularnewline
\midrule
LO MVP  & 0.0001 & 0.0333 & 1.0000 & 1.0000 & 0.0000 & 0.1056 & 26.5991 & 1.0000 & 1.0000 & 0.0000 \tabularnewline
LS MVP  & 0.0001 & 0.0333 & 1.0202 & 1.0101 & 0.0101 & 0.1048 & 26.4191 & 1.0000 & 1.0000 & 0.0000 \tabularnewline
LO C95  & 0.0001 & 0.0333 & 1.0000 & 1.0000 & 0.0000 & 0.1409 & 35.4975 & 1.0000 & 1.0000 & 0.0000 \tabularnewline
LS C95  & 0.0001 & 0.0333 & 1.0261 & 1.0130 & 0.0130 & 0.1408 & 35.4867 & 1.0000 & 1.0000 & 0.0000 \tabularnewline
LO C99  & 0.0001 & 0.0333 & 1.0000 & 1.0000 & 0.0000 & 0.1436 & 36.1841 & 1.0000 & 1.0000 & 0.0000 \tabularnewline
LS C99  & 0.0001 & 0.0333 & 1.0333 & 1.0167 & 0.0167 & 0.1439 & 36.2699 & 1.0000 & 1.0000 & 0.0000 \tabularnewline
LO TVP  & 0.0001 & 0.0333 & 1.0000 & 1.0000 & 0.0000 & 0.1558 & 39.2521 & 1.0000 & 1.0000 & 0.0000 \tabularnewline
LS TVP  & 0.0001 & 0.0333 & 1.0000 & 1.0000 & 0.0000 & 0.1556 & 39.2148 & 1.0000 & 1.0000 & 0.0000 \tabularnewline
LO TC95 & 0.0001 & 0.0333 & 1.0000 & 1.0000 & 0.0000 & 0.0019 & 0.4901 & 1.0000 & 1.0000 & 0.0000 \tabularnewline
LS TC95 & 0.0001 & 0.0333 & 1.0000 & 1.0000 & 0.0000 & 0.0020 & 0.5156 & 1.0000 & 1.0000 & 0.0000 \tabularnewline
LO TC99 & 0.0001 & 0.0333 & 1.0000 & 1.0000 & 0.0000 & 0.0021 & 0.5205 & 1.0000 & 1.0000 & 0.0000 \tabularnewline
LS TC99 & 0.0001 & 0.0333 & 1.0000 & 1.0000 & 0.0000 & 0.0019 & 0.4785 & 1.0000 & 1.0000 & 0.0000 \tabularnewline
\bottomrule
\end{tabular}%
}
\vspace{2pt}

\noindent{\footnotesize Notes: Daily turnover is computed as $0.5\sum_i |w_{i,t}-w_{i,t-1}|$. Annualized turnover equals average daily turnover multiplied by 252. Gross exposure is the sum of absolute portfolio weights. Short exposure is reported in absolute value.}
\end{table}

Tables~\ref{tab:tc_hist} and~\ref{tab:tc_dyn} report transaction-cost robustness for the historical and dynamic optimized portfolios. Transaction costs are applied proportionally to realized turnover. The no-cost case corresponds to 0 basis points per turnover, while the high-cost case applies 50 basis points per turnover. These tables show whether the reward-to-risk performance of each strategy is robust to implementation frictions.

\begin{table}[h!]
\centering
\caption{Transaction-cost robustness for historical optimized portfolios.}
\label{tab:tc_hist}
\scriptsize
\resizebox{\textwidth}{!}{%
\begin{tabular}{lrrrrrrr}
\toprule
& & \multicolumn{3}{c}{No transaction cost} & \multicolumn{3}{c}{50 bps per turnover} \tabularnewline
\cmidrule(lr){3-5}
\cmidrule(lr){6-8}
Strategy & Avg. Daily Turnover
& Sharpe & Calmar & STARR$_{0.95}$
& Sharpe & Calmar & STARR$_{0.95}$ \tabularnewline
\midrule
\multicolumn{8}{l}{\textbf{Minimum-risk portfolios}} \tabularnewline
\midrule
LO MVP  & 0.0001 & 0.5608 & 0.4524 & 4.7127 & 0.5600 & 0.4515 & 4.7072 \tabularnewline
LS MVP  & 0.0001 & 0.5594 & 0.4508 & 4.7040 & 0.5586 & 0.4498 & 4.6985 \tabularnewline
LO C95  & 0.0001 & 0.5476 & 0.4407 & 4.6200 & 0.5468 & 0.4397 & 4.6145 \tabularnewline
LS C95  & 0.0001 & 0.5453 & 0.4389 & 4.6028 & 0.5445 & 0.4379 & 4.5972 \tabularnewline
LO C99  & 0.0001 & 0.4672 & 0.3370 & 4.0529 & 0.4664 & 0.3363 & 4.0475 \tabularnewline
LS C99  & 0.0001 & 0.4622 & 0.3313 & 4.0189 & 0.4614 & 0.3306 & 4.0135 \tabularnewline
\midrule
\multicolumn{8}{l}{\textbf{Passive benchmark}} \tabularnewline
\midrule
BHP     & 0.0000 & 0.4453 & 0.3159 & 3.7256 & 0.4453 & 0.3159 & 3.7256 \tabularnewline
\midrule
\multicolumn{8}{l}{\textbf{Tangent portfolios}} \tabularnewline
\midrule
LO TVP  & 0.0001 & 0.4250 & 0.3210 & 3.7448 & 0.4243 & 0.3203 & 3.7395 \tabularnewline
LS TVP  & 0.0001 & 0.4250 & 0.3210 & 3.7448 & 0.4243 & 0.3203 & 3.7395 \tabularnewline
LO TC95 & 0.0001 & 0.4249 & 0.3215 & 3.7432 & 0.4241 & 0.3208 & 3.7379 \tabularnewline
LS TC95 & 0.0001 & 0.4249 & 0.3215 & 3.7432 & 0.4241 & 0.3208 & 3.7379 \tabularnewline
LO TC99 & 0.0001 & 0.4131 & 0.3040 & 3.6697 & 0.4124 & 0.3033 & 3.6644 \tabularnewline
LS TC99 & 0.0001 & 0.4131 & 0.3040 & 3.6697 & 0.4124 & 0.3033 & 3.6644 \tabularnewline
\bottomrule
\end{tabular}%
}
\vspace{2pt}
\noindent{\footnotesize Notes: Transaction costs are applied proportionally to realized portfolio turnover. The no-cost case corresponds to 0 basis points per turnover, while the high-cost case applies 50 basis points per turnover.}
\end{table}

\begin{table}[h!]
\centering
\caption{Transaction-cost robustness for dynamically optimized portfolios.}
\label{tab:tc_dyn}
\scriptsize
\resizebox{\textwidth}{!}{%
\begin{tabular}{lrrrrrrr}
\toprule
& & \multicolumn{3}{c}{No transaction cost} & \multicolumn{3}{c}{50 bps per turnover} \tabularnewline
\cmidrule(lr){3-5}
\cmidrule(lr){6-8}
Strategy & Avg. Daily Turnover
& Sharpe & Calmar & STARR$_{0.95}$
& Sharpe & Calmar & STARR$_{0.95}$ \tabularnewline
\midrule
\multicolumn{8}{l}{\textbf{Minimum-risk portfolios}} \tabularnewline
\midrule
LO MVP  & 0.1056 & 0.7255 & 0.5378 & 6.5516 & -0.5731 & -0.1334 & -2.5556 \tabularnewline
LS MVP  & 0.1048 & 0.7104 & 0.5376 & 6.4408 & -0.5762 & -0.1348 & -2.5799 \tabularnewline
LO C95  & 0.1409 & 0.8625 & 0.5234 & 7.2764 & -0.8035 & -0.2026 & -4.0401 \tabularnewline
LS C95  & 0.1408 & 0.7674 & 0.4699 & 6.6463 & -0.8824 & -0.2168 & -4.5997 \tabularnewline
LO C99  & 0.1436 & 0.8625 & 0.5498 & 7.2944 & -0.8267 & -0.2038 & -4.2110 \tabularnewline
LS C99  & 0.1439 & 0.7846 & 0.5007 & 6.8014 & -0.9059 & -0.2207 & -4.7692 \tabularnewline
\midrule
\multicolumn{8}{l}{\textbf{Passive benchmark}} \tabularnewline
\midrule
BHP     & 0.0000 & 0.3297 & 0.2464 & 3.0306 & 0.3297 & 0.2464 & 3.0306 \tabularnewline
\midrule
\multicolumn{8}{l}{\textbf{Tangent portfolios}} \tabularnewline
\midrule
LO TVP  & 0.1558 & -0.6571 & -0.2862 & -3.6284 & -1.2079 & -0.4200 & -6.8693 \tabularnewline
LS TVP  & 0.1556 & -0.6322 & -0.2835 & -3.4322 & -1.1766 & -0.4170 & -6.5934 \tabularnewline
LO TC95 & 0.0019 & 0.7979 & 0.5671 & 6.4383 & 0.7808 & 0.5524 & 6.3188 \tabularnewline
LS TC95 & 0.0020 & 0.7845 & 0.5472 & 6.3839 & 0.7663 & 0.5321 & 6.2562 \tabularnewline
LO TC99 & 0.0021 & 0.7599 & 0.5463 & 6.1935 & 0.7409 & 0.5301 & 6.0606 \tabularnewline
LS TC99 & 0.0019 & 0.7767 & 0.5297 & 6.2904 & 0.7596 & 0.5159 & 6.1705 \tabularnewline
\bottomrule
\end{tabular}%
}

\vspace{2pt}
\noindent{\footnotesize Notes: Transaction costs are applied proportionally to realized portfolio turnover. The no-cost case corresponds to 0 basis points per turnover, while the high-cost case applies 50 basis points per turnover.}
\end{table}

The historical transaction-cost results in Table~\ref{tab:tc_hist} are nearly unchanged between the no-cost and high-cost cases because historical turnover is extremely low. In contrast, Table~\ref{tab:tc_dyn} shows that several dynamically optimized portfolios are highly sensitive to transaction costs. Dynamic MVP, C95, C99, and TVP strategies have substantially higher turnover, so their performance deteriorates sharply when the 50 basis point cost is imposed. The dynamic TC95 and TC99 strategies are much more robust because their realized turnover is close to that of the historical strategies. These results show that the practical value of dynamic optimization depends not only on gross performance but also on whether improved performance can be achieved without excessive rebalancing.

\begin{table}[h!]
\centering
\caption{Historical and dynamic average absolute sector allocations.}
\label{tab:sector_abs_allocations}
\resizebox{\textwidth}{!}{%
\begin{tabular}{lrrrrrrrr}
\toprule
& \multicolumn{4}{c}{Historical}
& \multicolumn{4}{c}{Dynamic} \tabularnewline
\cmidrule(lr){2-5}
\cmidrule(lr){6-9}
Strategy
& Agriculture & Energy & Metals & Broad
& Agriculture & Energy & Metals & Broad \tabularnewline
\midrule
LO MVP  & 0.1103 & 0.2002 & 0.3895 & 0.3000 & 0.3860 & 0.0365 & 0.4503 & 0.1272 \tabularnewline
LS MVP  & 0.1103 & 0.2202 & 0.3897 & 0.3000 & 0.3863 & 0.0367 & 0.4505 & 0.1265 \tabularnewline
LO C95  & 0.1116 & 0.2000 & 0.3884 & 0.3000 & 0.3866 & 0.0487 & 0.4458 & 0.1189 \tabularnewline
LS C95  & 0.1116 & 0.2261 & 0.3884 & 0.3000 & 0.3898 & 0.0499 & 0.4453 & 0.1150 \tabularnewline
LO C99  & 0.1397 & 0.2047 & 0.3556 & 0.3001 & 0.3726 & 0.0493 & 0.4558 & 0.1224 \tabularnewline
LS C99  & 0.1397 & 0.2380 & 0.3556 & 0.3000 & 0.3751 & 0.0494 & 0.4562 & 0.1193 \tabularnewline
LO TVP  & 0.1065 & 0.2527 & 0.3408 & 0.3000 & 0.0365 & 0.5530 & 0.3673 & 0.0433 \tabularnewline
LS TVP  & 0.1065 & 0.2527 & 0.3408 & 0.3000 & 0.0381 & 0.5641 & 0.3554 & 0.0424 \tabularnewline
LO TC95 & 0.1077 & 0.2516 & 0.3407 & 0.3000 & 0.3751 & 0.1384 & 0.3771 & 0.1094 \tabularnewline
LS TC95 & 0.1077 & 0.2516 & 0.3407 & 0.3000 & 0.3641 & 0.1268 & 0.3783 & 0.1308 \tabularnewline
LO TC99 & 0.1101 & 0.2454 & 0.3445 & 0.3000 & 0.3565 & 0.1376 & 0.3750 & 0.1309 \tabularnewline
LS TC99 & 0.1101 & 0.2454 & 0.3445 & 0.3000 & 0.3277 & 0.1269 & 0.4018 & 0.1436 \tabularnewline
\bottomrule
\end{tabular}%
}
\vspace{2pt}

\noindent{\footnotesize Notes: Entries are average absolute portfolio weights aggregated by commodity category. For long-only strategies, absolute and net sector allocations coincide.}
\end{table}

Table~\ref{tab:sector_abs_allocations} reports average absolute portfolio weights aggregated by commodity category. The historical portfolios allocate exposure across agriculture, energy, metals, and broad commodity index funds, with broad commodity funds accounting for about 30\% of average exposure across most strategies. In the dynamic strategies, sector allocation varies more strongly across objectives. Dynamic MVP and CVaR minimum-risk portfolios allocate mainly to agriculture and metals, while dynamic TVP portfolios allocate more heavily to energy and metals. This sectoral reallocation helps explain why the dynamic tangent portfolios display weaker and more volatile performance in the main results.

Table~\ref{tab:top_etf_exposures} reports the three largest average ETF exposures for each strategy. The historical minimum-risk portfolios are driven mainly by gold-related and diversified agricultural exposures, such as GLDM, GLD, IAU, and DBA. The historical tangent portfolios place greater weight on UGA, CPER, CORN, SLV, and WEAT. In the dynamic results, MVP and CVaR minimum-risk portfolios are dominated by DBA and GLDM, while the TVP portfolios have large exposures to UGA, UNG, PALL, and SLV. This pattern is consistent with the main dynamic results, where the dynamic TVP portfolios are more fragile than the minimum-risk and CVaR-based allocations.

\begin{table}[h!]
\centering
\caption{Historical and dynamic largest average ETF exposures by strategy.}
\label{tab:top_etf_exposures}
\resizebox{\textwidth}{!}{%
\begin{tabular}{lllllll}
\toprule
& \multicolumn{3}{c}{Historical}
& \multicolumn{3}{c}{Dynamic} \tabularnewline
\cmidrule(lr){2-4}
\cmidrule(lr){5-7}
Strategy
& Largest & Second & Third
& Largest & Second & Third \tabularnewline
\midrule
LO MVP  & GLDM (0.0893) & DBA (0.0434) & GLD (0.0337) & DBA (0.3477) & GLDM (0.2695) & GLD (0.0762) \tabularnewline
LS MVP  & GLDM (0.0893) & DBA (0.0433) & GLD (0.0337) & DBA (0.3478) & GLDM (0.2720) & GLD (0.0729) \tabularnewline
LO C95  & GLDM (0.0792) & DBA (0.0447) & GLD (0.0426) & DBA (0.3378) & GLDM (0.2607) & TAGS (0.0847) \tabularnewline
LS C95  & GLDM (0.0792) & DBA (0.0447) & GLD (0.0426) & DBA (0.3385) & GLDM (0.2656) & TAGS (0.0827) \tabularnewline
LO C99  & WEAT (0.0682) & IAU (0.0558) & CORN (0.0371) & DBA (0.3148) & GLDM (0.2605) & TAGS (0.0963) \tabularnewline
LS C99  & WEAT (0.0682) & IAU (0.0558) & CORN (0.0371) & DBA (0.3143) & GLDM (0.2583) & TAGS (0.0943) \tabularnewline
LO TVP  & UGA (0.0522) & CPER (0.0408) & CORN (0.0386) & UGA (0.2350) & PALL (0.1507) & SLV (0.1357) \tabularnewline
LS TVP  & UGA (0.0522) & CPER (0.0408) & CORN (0.0386) & UGA (0.2163) & UNG (0.1451) & SLV (0.1356) \tabularnewline
LO TC95 & UGA (0.0516) & CPER (0.0410) & CORN (0.0394) & DBA (0.3352) & GLDM (0.1751) & IAU (0.0793) \tabularnewline
LS TC95 & UGA (0.0516) & CPER (0.0410) & CORN (0.0394) & DBA (0.3141) & GLDM (0.2240) & TAGS (0.0465) \tabularnewline
LO TC99 & UGA (0.0484) & SLV (0.0403) & WEAT (0.0399) & DBA (0.3222) & GLDM (0.2159) & TAGS (0.0471) \tabularnewline
LS TC99 & UGA (0.0484) & SLV (0.0403) & WEAT (0.0399) & DBA (0.2985) & GLDM (0.1549) & IAU (0.0977) \tabularnewline
\bottomrule
\end{tabular}%
}
\vspace{2pt}

\noindent{\footnotesize Notes: Entries report the three largest average ETF weights by absolute exposure. Parentheses contain average signed portfolio weights.}
\end{table}

\bibliographystyle{apalike}
\bibliography{ref}

\end{document}